\documentclass[iop,revtex4]{emulateapj}
\slugcomment{{\sc Accepted to ApJ:} January 22, 2013}
\usepackage{times}
\usepackage{graphicx}
\usepackage{amstext}
\usepackage{amsmath}
\usepackage{xspace}
\usepackage{color}
\usepackage{rotating}
\usepackage{txfonts}
\usepackage{booktabs}
\usepackage{url}
\usepackage{threeparttable}
\usepackage{lscape}
\bibpunct{(}{)}{;}{a}{}{,}

\newcommand{\Msun}{\ensuremath{M_\odot}\xspace}
 
\newcommand{\Teffsun}{\ensuremath{{T_{\text{eff}\,\odot}}}\xspace}
\newcommand{\loggsun}{\ensuremath{{\log g}_\odot}\xspace}
\newcommand{\Teff}{\ensuremath{{T_{\text{eff}}}}\xspace}
\newcommand{\logg}{\ensuremath{{\log g}}\xspace}
\newcommand{\loggf}{\ensuremath{{\log gf}}\xspace}
\newcommand{\vt}{\ensuremath{{v_t}}\xspace}
\graphicspath{
{./plots/}
}
\begin{document}

\title{Atmospheric composition of weak G band stars: CNO and Li abundances}

\author{Jens Adamczak and David L. Lambert}
\email{adamczak@astro.as.utexas.edu}
\affil{McDonald Observatory, The University of Texas, Austin, Texas, 78712, USA}

\begin{abstract}
We determined the chemical composition of a large sample of weak G band stars -- a rare class of G
and K giants of intermediate mass with unusual abundances of C, N, and Li. We have observed 24 weak
G band stars with the 2.7\,m Harlan J. Smith Telescope at the McDonald Observatory and derived
spectroscopic abundances for C, N, O, and Li, as well as for selected elements from Na - Eu. The
results show that the atmospheres of weak G band stars are highly contaminated with CN-cycle
products. The C underabundance is about a factor of 20 larger than for normal giants and the
$^{12}$C/$^{13}$C ratio approaches the CN-cycle equilibrium value. In addition to the striking
CN-cycle signature the strong N overabundance may indicate the presence of partially ON-cycled
material in the atmospheres of the weak G band stars. The exact mechanism responsible for the
transport of the elements to the surface has yet to be identified but could be induced by rapid
rotation of the main sequence progenitors of the stars. The unusually high Li abundances in some of
the stars are an indicator for Li production by the Cameron-Fowler mechanism. A quantitative
prediction of a weak G band star's Li abundance is complicated by the strong temperature sensitivity
of the mechanism and its participants. In addition to the unusual abundances of CN-cycle elements
and Li we find an overabundance of Na that is in accordance with the NeNa chain running in parallel
with the CN-cycle.  Apart from these peculiarities the element abundances in a weak G band star's
atmosphere are consistent with those of normal giants.  
\end{abstract}


\section{Introduction}
Theoretical and observational investigations of solar metallicity G and K giants have led to broad
agreement about the changes in surface composition arising from the first dredge-up initiated by the
deep convective envelope of the giant. Principal effects are a drastic lowering of the Li abundance,
a reduction of the $^{12}$C abundance, a lowering of the $^{12}$C/$^{13}$C ratio and an increase of
the $^{14}$N abundance with a preservation of the sum of $^{12}$C, $^{13}$C, and $^{14}$N nuclides. 

Rare examples of G and K giants exhibiting startling departures from the broad agreement shared by
the vast majority of the stars highlight the fact that facets of the evolution of these stars from
birth to the giant branch have eluded theoretical understanding.  Weak G band stars (here, wk Gb
stars) are just such a class of exceptional stars. The discovery of the first of these  stars
showing weak CH molecular lines in the Fraunhofer G band was made by
\nocite{1951ApJ...113..304B}{Bidelman} (1951) who examined low-resolution spectra of HR\,885 and
remarked `the spectrum is extraordinarily peculiar, with the line spectrum matching fairly well G5
III but with no trace of CN or CH absorption'. A quantitative analysis from coud\'{e} spectra was
undertaken by \nocite{1958ApJ...127..172G}{Greenstein} \& {Keenan} (1958)  who not only found both
CH and CN deficient by about 2 dex but uncovered a second wk Gb star -- HR\,6791 -- with less
extreme CH and CN deficiencies.

\nocite{1973AJ.....78..687B}{Bidelman} \& {MacConnell} (1973) laid the groundwork for an extensive
abundance analysis of wk Gb stars with a search of objective prism plates aimed at finding `the
brighter stars of astrophysical interest' which provided a list of 34 stars with `weak or no G
band'.  The survey forming the basis for the list was focused on the southern hemisphere and, hence,
did not include either HR\,885 or HR\,6791. Although the 1973 paper has been followed by several
papers discussing aspects of quantitative abundance analyses of wk Gb stars, there has, apart from
determinations of the Li abundance, been no comprehensive analysis of elements and isotopes (i.e.,
C, N, and O in particular) which should be useful diagnostics of the causes of the weak G band
phenomenon. Notably, \nocite{2012A&A...538A..68P}{Palacios} {et~al.} (2012) in their compilation of
published abundances for 28 wk G band stars list Li abundances for 18 stars but C and N abundances
for just four stars and O abundances for only two stars ( taken from
\nocite{1978ApJ...222..585S}{Sneden} {et~al.} 1978, \nocite{1978ApJ...221..893C}{Cottrell} \&
{Norris} 1978, and \nocite{1980PhDT.........5D}{Day} 1980). 

The principal aim of our paper is to provide the first thorough abundance analyses of a complete
sample of the wk Gb stars accessible to us at the W.J. McDonald Observatory in Texas. Our focus is
on the determination of the Li,  C, N, and O abundances and the $^{12}$C/$^{13}$C ratio but other
elements from Na to Eu are also included.  Such data are considered by us to be a prerequisite to a
discussion of likely origins of this class of peculiar red giants. Of the variety of origins
proposed in the literature, none seems to account convincingly for the reported abundance data.

\section{Observations}\label{sect:observations} We observed 24 wk Gb stars with the 2.7\,m Harlan J.
Smith Telescope at the McDonald Observatory.  The telescope was equipped with the Robert G. Tull
Cross-Dispersed Echelle Spectrograph \nocite{1995PASP..107..251T}({Tull} {et~al.} 1995) and a
Tektronix 2048\,x\,2048 pixel CCD detector. A majority of the stars was observed with two different
setups of the Tull spectrograph. The pair when combined provide coverage of the inter-order echelle
gaps. The \textsl{blue} setup was centered on 5112\,\AA\xspace in order 68, while the \textsl{red}
setup is centered on 5178\,\AA\xspace (order 68). A few stars were observed in what we deem the
standard (std) setup centered on 5060\,\AA\xspace in order 69. For all setups a wavelength range of
3\,900--10\,000\,\AA\xspace was covered with a resolving power of $\text{R}\equiv \lambda/ \Delta
\lambda = 60\,000$. The observations are listed in Table \ref{tab:observations}. The spectra were
reduced in a multiple step procedure. First flat field and bias corrections were applied. Then the
echelle spectra were extracted and wavelength calibrated using ThAr-lamp comparison exposures taken
before or after the observations. The different spectral orders were co-added and continuum
normalized. All tasks were performed using the IRAF package \footnote{IRAF is distributed by the
National Optical Astronomy Observatories, which are operated by the Association of Universities for
Research in Astronomy, Inc., under cooperative agreement with the National Science Foundation.}. 

\tabletypesize{\small}
\begin{deluxetable}{lrrl}
\tablecaption{Observation log of analyzed spectra. The different setups are described in the
text.\label{tab:observations}} \tablewidth{0pt}
\tablehead{
  \colhead{Object}    & \colhead{Obs. date (UT time)}   & \colhead{$t_{\text exp}$[s]} &
  \colhead{setup}
}
\startdata
37\,Com             & Feb 19 2011 & $600$             & \textsl{red}  \\
                    & Feb 20 2011 & $600$             & \textsl{blue} \\
BD\,+5$^{\circ}$593 & Feb 19 2011 & $3\times 1\,800$  & \textsl{red}  \\
                    & Feb 20 2011 & $3\times 1\,800$  & \textsl{blue} \\
HD\,18636           & Nov 15 2011 & $3\times 1\,200$  & \textsl{red}  \\
                    & Nov 16 2011 & $3\times 1\,200$  & \textsl{blue} \\
HD\,28932           & Nov 15 2011 & $4\times 1\,200$  & \textsl{red}  \\
                    & Nov 16 2011 & $4\times 1\,200$  & \textsl{blue} \\
HD\,31869           & Nov 16 2011 & $4\times 1\,200$  & \textsl{blue} \\
HD\,40402           & Nov 16 2011 & $3\times 1\,200$  & \textsl{blue} \\
                    & Nov 17 2011 & $2\times 1\,800$  & \textsl{std}  \\
HD\,49960           & Nov 15 2011 & $3\times 1\,200$  & \textsl{red}  \\
                    & Nov 16 2011 & $4\times 1\,200$  & \textsl{blue} \\
HD\,67728           & Feb 19 2011 & $2\times 1\,800$  & \textsl{red}  \\
                    & Feb 20 2011 & $2\times 1\,800$  & \textsl{blue} \\
HD\,78146           & Feb 19 2011 & $3\times 1\,800$  & \textsl{red}  \\
                    & Feb 20 2011 & $3\times 1\,800$  & \textsl{blue} \\
HD\,82595           & Feb 19 2011 & $2\times 1\,800$  & \textsl{red}  \\
                    & Feb 20 2011 & $2\times 1\,800$  & \textsl{blue} \\
HD\,94956           & Feb 19 2011 & $2\times 1\,800$  & \textsl{red}  \\
                    & Feb 20 2011 & $2\times 1\,800$  & \textsl{blue} \\
HD\,120170          & Feb 19 2011 & $3\times 1\,800$  & \textsl{red}  \\
                    & Feb 20 2011 & $3\times 1\,800$  & \textsl{blue} \\
HD\,120171          & Feb 19 2011 & $3\times 1\,800$  & \textsl{red}  \\
                    & Feb 20 2011 & $3\times 1\,800$  & \textsl{blue} \\
HD\,132776          & Mar 07 2012 & $2\times 1\,800$  & \textsl{blue} \\
                    & Mar 08 2012 & $2\times 1\,800$  & \textsl{red}  \\
HD\,146116          & Mar 07 2012 & $2\times 1\,200$  & \textsl{blue} \\
                    & Mar 08 2012 & $3\times 1\,800$  & \textsl{red}  \\
HD\,188028          & Sep 24 2012 & $1\,200 + 1\,800$ & \textsl{blue} \\
                    & Sep 25 2012 & $1\,200 + 1\,800$ & \textsl{red}  \\         
HD\,204046          & Nov 16 2011 & $3\times 1\,200$  & \textsl{blue} \\
                    & Nov 17 2011 & $3\times 1\,200$  & \textsl{red}  \\
HD\,207774          & Nov 15 2011 & $4\times 1\,200$  & \textsl{red}  \\
                    & Nov 16 2011 & $4\times 1\,200$  & \textsl{blue} \\
HR\,885             & Feb 21 2011 & $900$             & \textsl{std}  \\
HR\,1023            & Feb 21 2011 & $1\,200$          & \textsl{std}  \\
HR\,1299            & Nov 15 2011 & $2\times 1\,200$  & \textsl{red}  \\
                    & Nov 16 2011 & $2\times 1\,200$  & \textsl{blue} \\
HR\,6757            & Feb 19 2011 & $2\times 1\,200$  & \textsl{red}  \\
                    & Feb 20 2011 & $2\times 1\,200$  & \textsl{blue} \\
HR\,6766            & May 14 2011 & $360$             & \textsl{red}  \\
                    & May 14 2011 & $300$             & \textsl{blue} \\
HR\,6791            & May 14 2011 & $360$             & \textsl{red}  \\
                    & May 14 2011 & $360$             & \textsl{blue} 
\enddata
\end{deluxetable}
 
\section{Parameter determination}\label{sect:parameters}         
\subsection{Spectroscopic}
The parameters of the program stars were derived by spectroscopic means using Fe\,{\sc i} and
Fe\,{\sc ii} lines. The line selection process included the comparison of different line lists and
atomic data. The original line list composed for this purpose contained about 280 Fe\,{\sc i} and 25
Fe\,{\sc ii} lines that we identified in the wk Gb spectra by comparison with the solar atlas.
Values for the \loggf values of these lines were taken from NIST \footnote{\textsl{National
Institute of Standard and Technology} atomic spectra database (\url{http://physics.nist.gov/})}, the
Kurucz database \nocite{1995KurCD..23.....K}({Kurucz} \& {Bell} 1995), and
\nocite{2009A&A...497..611M}{Mel{\'e}ndez} \& {Barbuy} (2009). The line equivalent widths (EWs) were
measured with an \textsl{Automatic Routine for line Equivalent widths in stellar Spectra}
(\textsl{ARES}, \nocite{2007A&A...469..783S}{Sousa} {et~al.} 2007). Lines not found, or not
identified as single lines by \textsl{ARES}, were removed from the list. As a next step, the
measured EWs were plotted against the strength of the lines, i.e.  roughly the difference of \loggf
and excitation potential $\chi$, in order to identify and remove lines, which deviate too strongly
from the linear trend.  The resulting line list  was then used as input for the spectral synthesis
code \textsl{MOOG} \nocite{1973PhDT.......180S}({Sneden} 1973) for an abundance analysis. We made
use of input model atmospheres, which were interpolated from a grid of models from
\nocite{2004astro.ph..5087C}{Castelli} \& {Kurucz} (2004). Assuming a one-dimensional,
plane-parallel atmosphere in local thermodynamic equilibrium (LTE), the abundances of the computed
lines were force-fitted to match the measured EWs.  The derived abundances were plotted vs. the
excitation potential of the lines and any trend was flattened by adjusting the effective temperature
(\Teff) of the model atmosphere. Simultaneously, the microturbulence was derived by reducing the
trend in the plot of Fe abundances and reduced equivalent width (EW divided by wavelength). The
surface gravity was adjusted until abundances derived by Fe\,{\sc i} and Fe\,{\sc ii} lines matched
within 0.12 dex, which for most of the stars corresponds to the standard deviation of the Fe\,{\sc
ii} abundance. Since all parameters are to some extent interdependent, usually several iterations
are needed. In order to minimize the error of the \Teff determination, the line list was iteratively
cleaned of Fe\,{\sc i} and Fe\,{\sc ii} lines that showed significant deviations in their abundances
compared to the derived average abundance of their species. This was done until the resulting
$\sigma$ for the abundance determination was reduced to $\approx$ 0.05.  NLTE effects that could
affect the determination of Fe abundances and stellar parameters are negligible for G and K stars in
our metallicity range (\nocite{2011JPhCS.328a2002B}{Bergemann} {et~al.} 2011,
\nocite{2011A&A...528A..87M}{Mashonkina} {et~al.} 2011).  We repeated the procedure for the line
lists presented in \nocite{2011ApJ...743..135R}{Ram{\'{\i}}rez} \& {Allende Prieto} (2011) and
\nocite{2008PASJ...60..781T}{Takeda}, {Sato}, \&  {Murata} (2008). The former consists of 37
Fe\,{\sc i} and 9 Fe\,{\sc ii} lines with \loggf values from different sources (e.g the Oxford group
and \nocite{2002ApJ...573L.137A}{Allende Prieto}, {Lambert}, \&  {Asplund} 2002, most references can
be found in \nocite{1996ApJS..103..183L}{Lambert} {et~al.} 1996) and was compiled for the
determination of the iron abundance of Arcturus. The latter includes 124 Fe\,{\sc i} and 11 Fe\,{\sc
ii} lines with excitation potential and \loggf values from \nocite{1999A&A...347..348G}{Grevesse} \&
{Sauval} (1999), and was used for the determination of stellar parameters of late-G giants.

Effective temperatures derived with the line list from \nocite{2011ApJ...743..135R}{Ram{\'{\i}}rez}
\& {Allende Prieto} (2011) agree with the temperatures that were derived with our original large
line list. The scatter is less than $\pm 100\,\text{K}$.  The logarithms of the surface gravities
agree within $\pm 0.3\,\text{dex}$. The microturbulence derived with the
\nocite{2011ApJ...743..135R}{Ram{\'{\i}}rez} \& {Allende Prieto} (2011) line list is on average
higher by about $0.5\,\text{km s}^{-1}$, the metallicity lower by about 0.15\,dex. 

The temperatures derived with the \nocite{2008PASJ...60..781T}{Takeda} {et~al.} (2008) line list are
on average lower than the ones derived with the original line list by about 150\,K, the \logg values
are lower by about 0.3\,dex, microturbulences are higher by $0.3\,\text{km s}^{-1}$, and
metallicities are smaller by about 0.25\,dex.

The atomic data for the lines that our original list and
\nocite{2011ApJ...743..135R}{Ram{\'{\i}}rez} \& {Allende Prieto} (2011) have in common is almost
identical, with differences in the \loggf values of $\Delta \loggf = -0.031\pm 0.076$. The small
discrepancy in the derived parameters is therefore probably a result of the line selection itself. A
higher number of lines of a certain excitation potential could have statistical influence on the
observed trend of abundance with $\chi$. To test the influence of the number of lines included in a
line list, we randomly chose a subset of lines from our original list and repeated the parameter
determination. The results, however, did not vary significantly. We decided to compile a final line
list out of the line list from \nocite{2008PASJ...60..781T}{Takeda} {et~al.} (2008) and
\nocite{2011ApJ...743..135R}{Ram{\'{\i}}rez} \& {Allende Prieto} (2011) and include the lines of
\nocite{2009A&A...497..611M}{Mel{\'e}ndez} \& {Barbuy} (2009) for additional Fe\,{\sc ii} line data.
The Kurucz loggf values are computed and might therefore be less accurate. The line lists of
Takeda et al. and Ram{\'{\i}}rez \& Allende Prieto are compiled and tested for giant stars and include
reliable loggf values. Hence the choice for the final line list.
The spectroscopic (sp.) parameters derived with this line list can be found in Table
\ref{tab:parameters}. 

The uncertainties for the temperatures can be
estimated from different sources. Since we typically derived two temperatures for every star,
corresponding to the different observing setups (see section \ref{sect:observations}), we take the
difference between both derived temperatures as a statistical error. This error is $\approx
30\,\text{K}$. In cases where only one setup was available we assume this error to be 50\,K. For
each setup we determined an additional error by varying the temperature and investigating the impact
on the slope of the trend between Fe\,{\sc i} abundances and excitation potential. Varying the
temperature by more than 50\,K results in a difference in the Fe abundances of lines with low and
high excitation potential that is larger than the $1 \sigma$ error derived for the Fe abundance with
no trend. We take this as our instrinsic error. Adding both errors results in a mean error for the
spectroscopic temperatures of 90\,K (see Table  \ref{tab:parameters}).

For the spectroscopic gravities we estimate a mean error of 0.18\,dex from the different gravities
derived for each star with the different setups. The metallicities in Table \ref{tab:parameters} are
calculated as the mean value of Fe\,{\sc i} and Fe\,{\sc ii} abundances, minus the solar Fe
abundance of \nocite{2009ARA&A..47..481A}{Asplund} {et~al.} (2009), $\log \epsilon(\text{Fe})=7.50$
\footnote{$\log \epsilon (\text{X}) = \log(N_\text{X}/N_\text{H}) + 12$}, of all setups that were
available for a star. The quoted errors are derived by taking into account the standard deviation of
these abundances and the errors for the individual abundance determinations. In cases where only one
setup was available we assume an error of 0.1 for the missing abundances.
\begin{figure}
\centering
\includegraphics[angle=-90]{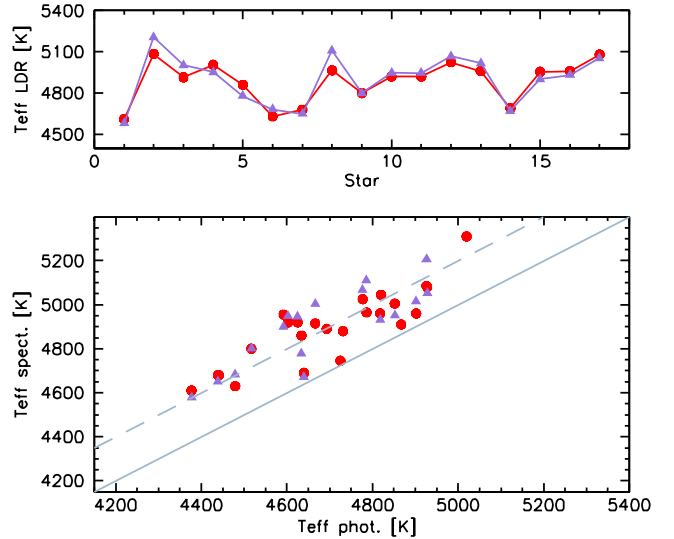}
\caption{Top panel: Comparison of \Teff ranking for spectroscopic (circles) and LDR temperatures
(triangles). Lower panel: Comparison of photometric temperatures lines with spectroscopic \Teff
determined using Fe\,{\sc i} lines (circles) and \Teff derived with the line-depth ratio method
(triangles). The spectroscopic temperatures are consistently higher than the photometric ones. The
dashed line indicates an offset of 200\,K from the 1:1 ratio (solid line). \label{fig:f1}}  
\end{figure}

\subsection{Line Depth Ratios}
To check that our line list gives accurate results for the \Teff of our objects we employed
additional methods to determine their temperatures. As described in
\nocite{2001PASP..113..723G}{Gray} \& {Brown} (2001), line depth ratios (LDRs) of certain elements
are excellent indicators of stellar temperatures for G and K giants. Measuring the depth of
temperature sensitive V\,{\sc i} lines and comparing them to close-by Ni\,{\sc i}, Si\,{\sc i} and
Fe\,{\sc i} lines, provides a good tool to constrain the temperatures of the stars.
\nocite{2001PASP..113..723G}{Gray} \& {Brown} (2001) calibrated the LDRs against the (B-V) color
index and applied corrections for metallicity and absolute-magnitude variations. The (B-V) colors
were then converted into temperatures using the temperature scale of
\nocite{1999A&AS..134..523T}{Taylor} (1999). We use our spectroscopically derived metallicities and
parallaxes from the new \textsl{Hipparcos} reduction of \nocite{2007A&A...474..653V}{van Leeuwen}
(2007) for the metallicity and absolute-magnitude corrections. 

Our temperatures determined by the LDRs measured in the wk Gb stars are in excellent agreement with
our spectroscopically derived temperatures, the mean difference being 13\,K. It should, however, be
noted that the dependence on an original effective temperature scale limits the accuracy of the
determination of absolute temperatures derived by the LDR measurements. Nevertheless, the ranking of
the stars according to their relative temperatures is not affected and also agrees well with the
temperature sequence found for our spectroscopic temperatures (cf. Figure \ref{fig:f1}). For some
stars no LDR temperatures could be derived due to different reasons. For the stars with imprecisely
known parallaxes, no correction for absolute-magnitude variations could be made. For 37\,Com,
HD\,67728, and HR\,1023, strong line broadening caused difficulties for the LDR measurement. 
The LDR method depends on a low $v \sin i$ and these three stars are rapid rotators.
However, for all stars with measured LDRs the derived temperatures confirm our spectroscopic result
and justify our particular choice of line list. 
\begin{deluxetable*}{llrlrlllllrl}
\tablecaption{Atmospheric parameters of the program stars.\label{tab:parameters}}
\tablewidth{0pt}
\tablehead{
\colhead{Object}    &\colhead{\Teff sp.}     &\colhead{$\sigma$}  &\colhead{\Teff ph.}     &\colhead{$\sigma$}   &\colhead{\logg sp.}     &
\colhead{$\sigma$}  &\colhead{\logg ph.}     &\colhead{$\sigma$}  &\colhead{\vt sp.}       &\colhead{[Fe/H] sp.} &\colhead{$\sigma$}      \\
\colhead{}          &\colhead{[K]}           &\colhead{}          &\colhead{[K]}           &\colhead{}           &\colhead{[cm s$^{-2}$]} &         
\colhead{}          &\colhead{[cm s$^{-2}$]} &\colhead{}          &\colhead{[km s$^{-1}$]} &\colhead{}           &\colhead{}             
}
\startdata
37\,Com             & 4610      & 100     & 4377                  &  34     & 2.5           & 0.18    & 1.9           &  0.11   & 2.8           & -0.53      & 0.11    \\
BD\,+5$^{\circ}$593 & 5045      &  60     & 4820\tablenotemark{a} & 188     & 2.5           & 0.18    &               &         & 1.6           & -0.29      & 0.14    \\
HD\,18636           & 5085      &  80     & 4926                  &  57     & 2.7           & 0.18    & 2.8           &  0.17   & 1.6           & -0.16      & 0.13    \\
HD\,28932           & 4915      &  80     & 4666                  &  62     & 2.5           & 0.18    & 2.3           &  0.34   & 1.5           & -0.39      & 0.14    \\
HD\,31869           & 4800      & 100     & 4867                  &  82     & 1.8           & 0.18    & 3.2           &  3.33   & 1.6           & -0.47      & 0.20    \\
HD\,40402           & 5005      & 110     & 4852                  &  73     & 2.8           & 0.18    & 2.6           &  0.66   & 1.3           & -0.14      & 0.14    \\
HD\,49960           & 4860      & 130     & 4634                  &  81     & 2.4           & 0.18    & 2.0           &  0.65   & 1.7           & -0.25      & 0.16    \\
HD\,67728           & 4630      & 100     & 4479                  &  41     & 1.2           & 0.18    & 2.2           &  0.29   & 1.9           & -0.43      & 0.18    \\
HD\,78146           & 4680      &  90     & 4439                  &  75     & 2.1           & 0.18    & 1.8           &  1.04   & 1.6           & -0.06      & 0.12    \\
HD\,82595           & 4880      &  60     & 4731                  &  57     & 2.5           & 0.18    & 2.6           &  0.29   & 1.6           &  0.06      & 0.15    \\
HD\,94956           & 4965      &  60     & 4786                  &  69     & 2.5           & 0.18    & 2.7           &  0.41   & 1.7           & -0.19      & 0.13    \\
HD\,120170          & 4890      &  90     & 4693                  &  65     & 2.4           & 0.18    & 4.2           &  0.66   & 1.5           & -0.51      & 0.16    \\
HD\,120171          & 4745      & 120     & 4725                  &  82     & 2.3           & 0.18    &               &         & 1.5           & -0.42      & 0.15    \\
HD\,132776          & 4680      &  70     & 4441                  & 130     & 2.3           & 0.18    & 2.0           &  0.98   & 1.5           & -0.07      & 0.17    \\
HD\,146116          & 4800      &  90     & 4517                  &  70     & 2.1           & 0.18    & 1.8           &  0.54   & 1.7           & -0.39      & 0.14    \\
HD\,188028          & 4920      &  70     & 4603                  &  90     & 2.3           & 0.18    & 2.1           &  0.45   & 1.7           & -0.18      & 0.13    \\
HD\,204046          & 4920      &  90     & 4625                  & 158     & 2.5           & 0.18    & 2.4           &  0.81   & 1.7           & -0.05      & 0.14    \\
HD\,207774          & 5025      & 160     & 4777                  &  17     & 2.6           & 0.18    & 2.8           &  0.51   & 1.7           & -0.33      & 0.12    \\
HR\,885             & 4960      & 100     & 4902                  &  53     & 2.1           & 0.18    & 2.5           &  0.06   & 1.7           & -0.35      & 0.16    \\
HR\,1023            & 5310      & 100     & 5020                  &  33     & 1.6           & 0.18    & 2.2           &  0.25   & 2.5           & -0.22      & 0.18    \\
HR\,1299            & 4690      &  60     & 4640                  &  35     & 2.2           & 0.18    & 2.3           &  0.11   & 1.4           & -0.08      & 0.15    \\
HR\,6757            & 4955      &  60     & 4592                  &  97     & 2.5           & 0.18    & 2.4           &  0.11   & 1.5           & -0.08      & 0.11    \\
HR\,6766            & 4960      &  70     & 4818                  &  60     & 2.4           & 0.18    & 2.4           &  0.04   & 1.7           & -0.18      & 0.14    \\
HR\,6791            & 5080      &  90     & 4928                  &  61     & 2.6           & 0.18    & 2.5           &  0.08   & 1.5           &  0.00      & 0.15    
\enddata
\tablenotetext{a}{Only DDO photometric data was available.}
\end{deluxetable*}

\subsection{Photometric}
A multitude of different sources for photometric data and calibrations exist. One of the most often
applied is the calibration from \nocite{1999A&AS..140..261A}{Alonso}, {Arribas}, \&
{Mart{\'{\i}}nez-Roger} (1999). In order to derive precise and accurate photometric temperatures we
try to derive temperatures from as many colors as possible and use only quality photometric data. We
employ the General Catalogue of Photometric Data (GCPD, \nocite{1997A&AS..124..349M}{Mermilliod},
{Mermilliod}, \&  {Hauck} 1997) as well as ground-based V magnitudes included in the
\textsl{Hipparcos}-Tycho catalogue (ESA 1997). Infrared colors from the Two Micron All Sky Survey
(2MASS) were also used, if the observations were not saturated, and transformed into the TCS
(Telescopio Carlos S{\'a}nchez) system used by \nocite{1999A&AS..140..261A}{Alonso} {et~al.} (1999).
We also made use of the calibration of \nocite{2005ApJ...626..465R}{Ram{\'{\i}}rez} \&
{Mel{\'e}ndez} (2005). They offer calibrations for a lot of different photometric systems including
\textsl{UBV}, \textsl{uvby}, Vilnius, Geneva, and DDO in addition to the systems included in the
\nocite{1999A&AS..140..261A}{Alonso} {et~al.} (1999) calibrations. Also, the calibrations of
\nocite{2005ApJ...626..465R}{Ram{\'{\i}}rez} \& {Mel{\'e}ndez} (2005) avoid uncertainties due to
color-color transformations. For all calibrations we used our spectroscopically determined
metallicities.

The temperatures derived from the (B-V) color indices with the calibration of
\nocite{1999A&AS..140..261A}{Alonso} {et~al.} (1999) agree well with the corresponding temperatures
derived with the \nocite{2005ApJ...626..465R}{Ram{\'{\i}}rez} \& {Mel{\'e}ndez} (2005) calibrations,
the mean difference being lower than 10\,K. For the (V-K) colors differences are $\approx
20\,\text{K}$. These are the only indices that both calibration have in common. The photometric
temperatures derived from different systems show some scatter. It is possible that the weak G band
in our objects is affecting the photometric colors - directly, or indirectly. A direct effect is
most serious for colors from the DDO system \nocite{1968AJ.....73..313M}({McClure} \& {van den
Bergh} 1968). The temperatures derived from this system are higher than the ones derived from all
other systems.  \nocite{2005ApJ...626..465R}{Ram{\'{\i}}rez} \& {Mel{\'e}ndez} (2005) report a
strong metallicity dependence for the DDO colors, however, the main reason for the discrepancy is
that the giant sample that \nocite{2005ApJ...626..465R}{Ram{\'{\i}}rez} \& {Mel{\'e}ndez} (2005)
used to calibrate the Teff scale is dominated by stars with normal G-band. This leads to a lack of
absorption in the 42 filter and makes the star bluer and therefore hotter (Ram{\'{\i}}rez, priv.
comm.). Our final photometric temperatures are derived by averaging all available temperatures
without including those derived by the DDO colors whenever possible \footnote{In the cases of
BD\,+5$^{\circ}$593 only DDO photometric data was available.  In these cases we take the mean of the
temperatures derived from the C(42-45) and $\text{C}(42-48)\equiv \text{C}(42-45)+\text{C}(45-48)$
colors.}. They can be found in Table  \ref{tab:parameters} together with their errors, which are
simply the standard deviations of the mean temperatures. 

The photometric temperatures are consistently lower by 194\,K than the spectroscopic ones (see lower
panel of Figure \ref{fig:f1}). It is not clear where the offset between spectroscopic and
photometric temperatures originates. A similar comparison between photometric temperatures derived
with (B-V) colors and spectroscopic temperatures in \nocite{2008PASJ...60..781T}{Takeda} {et~al.}
(2008) shows a dispersion of comparable size but no systematic offset. Interestingly, in the paper
of \nocite{2007AJ....133.2464L}{Luck} \& {Heiter} (2007) on giants in the local region, the authors
find an offset of 98\,K between their spectroscopic temperatures derived with the newest MARCS
models and their photometrically derived temperatures. This offset increases by another 50\,K when
using the older MARCS75 model atmospheres.  \nocite{2010ApJ...710.1365B}{Baines} {et~al.} (2010)
present spectroscopic temperatures for 25 K giants. We determined photometric temperatures for their
objects and find that the spectroscopic temperatures are higher by about 80\,K. These comparisons
show that higher spectroscopic temperatures are not specific to the wk Gb stars. Their main
characteristic, the weak absorption at wavelengths around 4300\,\AA, might affect single colors only
(mainly B, and the 42 filter of the DDO system). However, the higher flux in these filters would
only increase the photometrically derived temperatures, and thus, minimize the difference between
the temperatures. 

It should be noted that the differences in the temperatures derived with the different methods
presented here are not critical to the analysis and results described in the following sections. The
differences in the derived abundances (see sections \ref{sect:carbon}-\ref{sect:otherelements}) are
subtle and do not affect the overall conclusions derived in these sections and discussed in
sections \ref{sect:discussion} and \ref{sect:conclusion}.
  
\subsection{Reddening, mass, and luminosity} 
We estimated the reddening of our stars similar to the procedure described in
\nocite{2007AJ....133.2464L}{Luck} \& {Heiter} (2007). First we calculate the reddening of our stars
with the EXTINCT code provided by \nocite{1997AJ....114.2043H}{Hakkila} {et~al.} (1997), using the
position of our objects and the distances calculated from the \textsl{Hipparcos} parallaxes as
input. Then we subtract the reddening calculated for a distance of 75\,pc accounting for the
reddening-free Local Bubble.

Most of our objects are not affected by reddening. Over 50\% of the sample stars have a reddening
lower than $E(B-V)=0.025$ which corresponds to a change in \Teff of $\approx +50\,\text{K}$ for the
(B-V) colors. The determination of the reddening in most of the other cases suffers from uncertain
parallaxes and large errors in the determined total visual extinction, sometimes far exceeding the
actual values.  We therefore continue under the assumption that reddening is negligible.

\begin{figure*}
\centering
\includegraphics[angle=-90]{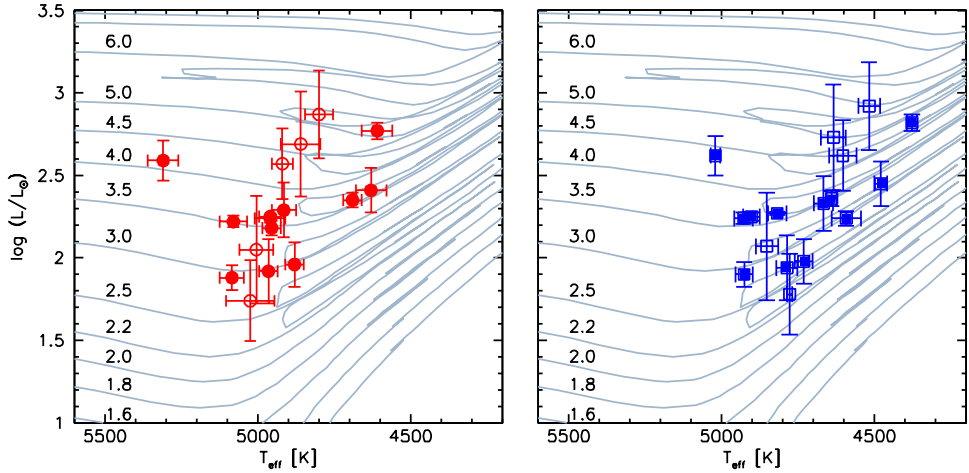}
\caption{Program stars with spectroscopically (left panel) and photometrically (right panel)
determined \Teff in the HRD. Filled symbols mark objects with parallax errors smaller than 50\%.
Evolutionary tracks, labeled with the respective stellar masses in \Msun, were calculated by
\protect\nocite{2008A&A...484..815B,2009A&A...508..355B}{Bertelli} {et~al.} (2008, 2009) for a solar
composition ($\text{Z}_\odot=0.017$, $\text{Y}_\odot=0.260$).  \label{fig:f2}}  
\end{figure*}

We calculated luminosities for the program stars with the \textsl{Hipparcos} parallaxes and the GCPD
visual magnitudes and their errors. In the cases where no V magnitude or error estimate was
available in the GCPD we take magnitudes and errors from the NOMAD catalogue
\footnote{http://www.nofs.navy.mil/nomad/}. We applied a bolometric correction according to
\nocite{1999A&AS..140..261A}{Alonso} {et~al.} (1999). Using both spectroscopic and photometric
temperatures we derived the position of the stars in the Hertzsprung-Russell Diagram (HRD). For some
of the stars the errors in the parallaxes are large. Figure \ref{fig:f2} shows the location of the
wk Gb stars in the HRD and evolutionary tracks from
\nocite{2008A&A...484..815B,2009A&A...508..355B}{Bertelli} {et~al.} (2008, 2009). The errors of the
parallaxes of some of our objects are large. We therefore only plot objects with parallax errors
smaller than 75\% of the actual parallax.  Depending on the set of temperatures used, the positions
of the objects lead to different conclusions.  While the hotter spectroscopic temperatures place the
objects at positions at the beginning of the giant branch ascent, the photometric temperatures
indicate a more advanced evolution. It is apparent that for the region in the HRD where the wk Gb
stars can be found, the exact placement of the stars depends critically on the temperature.

We determined the masses of the program stars by interpolating between the evolutionary tracks of
\nocite{2008A&A...484..815B,2009A&A...508..355B}{Bertelli} {et~al.} (2008, 2009). We do not use a
complete isochrone fitting technique and determine probability distribution functions for our masses
but rather rely on the evolutionary tracks for a solar composition ($\text{Z}_\odot=0.017$ and
$\text{Y}_\odot=0.260$).  Changing the metallicity does not siginifcantly affect the masses of our
objects.  We make a test calculation for this approach for Arcturus and its parameters given in
\nocite{2011ApJ...743..135R}{Ram{\'{\i}}rez} \& {Allende Prieto} (2011) and obtain
$M_\text{A}=1.47\pm0.08\,\Msun$ compared to their value of $1.08\pm0.06\,\Msun$ and justifies
our approach. The masses of the wk Gb stars are in the range $2.5-5\Msun$ with a mean mass of
$3.56\pm1.07\,\Msun$. 

With the masses of the stars we determined photometric surface gravities using our parameters and
the relation
\begin{align*}
\logg = 0.4(M_\text{bol}+M_{\text{bol}\,\odot}) + \loggsun +
4\log \left( \frac{\Teff}{\Teffsun} \right) + \log \left( \frac{M}{\Msun} \right),
\end{align*}
with $M_\text{bol}=M_\text{V}+BC$. We take $\Teffsun=5780\,\text{K}$, $\loggsun=4.44$, and
$M_{\text{bol}\,\odot}=4.75\,\text{mag}$, as the solar parameters. The photometrically derived
gravities are close to the spectroscopic ones, the mean difference being $0.13\pm0.57$. For some of
the stars however, large errors arise from uncertainties in parallaxes (for the calculation of
$M_\text{bol}$) and masses. For BD\,+5$^{\circ}$593 and HD\,120171, for example, no \logg could be
derived at all. We therefore rely exclusively on the spectroscopic gravities in the further analysis. Our derived
masses for the wk Gb stars can be found in Table \ref{tab:masses}. Note that we typically obtain
unsymmetrical errors for the masses and quoted the higher ones in the table.
\begin{deluxetable*}{lrrrrrrrrr}
\tablecaption{Distances and masses of the program stars.\label{tab:masses}}
\tablewidth{0pt}
\tablehead{
\colhead{Object}    &\colhead{plx}      &\colhead{$\sigma$} &\colhead{$\sigma$} &\colhead{d}    
                    &\colhead{$\sigma$} &\colhead{mass sp.} &\colhead{$\sigma$} &\colhead{mass ph.} &\colhead{$\sigma$}\\
\colhead{}          &\colhead{[mas]}    &\colhead{}         &\colhead{[\%]}    &\colhead{[pc]}     
                    &\colhead{}         &\colhead{[\Msun]}  &\colhead{}         &\colhead{[\Msun]}  &\colhead{}        
}
\startdata
37\,Com             &  4.94&  0.55  &  11.1  &  202 &   23   & 4.8      & 0.3    & 4.3      & 0.5   \\ 
HD\,18636           &  3.48&  0.57  &  16.4  &  287 &   47   & 2.9      & 0.3    & 2.9      & 0.3   \\
HD\,28932           &  1.94&  0.73  &  37.6  &  516 &  194   & 3.7      & 0.7    & 3.5      & 1.1   \\
HD\,31869           &  3.90& 14.70  & 376.9  &  256 &  967   & 1.7      & 8.1    & 1.9      & 8.0   \\
HD\,40402           &  1.87&  1.39  &  74.3  &  535 &  398   & 3.2      & 1.3    & 3.2      & 1.5   \\
HD\,49960           &  1.03&  0.76  &  73.8  &  971 &  716   & 4.7      & 1.8    & 4.7      & 2.1   \\
HD\,67728           &  2.16&  0.67  &  31.0  &  463 &  144   & 3.7      & 0.9    & 3.2      & 1.1   \\
HD\,78146           &  0.84&  0.99  & 117.9  & 1191 & 1403   & 5.1      & 3.2    & 4.8      & 4.9   \\
HD\,82595           &  2.58&  0.79  &  30.6  &  388 &  119   & 3.0      & 0.6    & 2.6      & 0.7   \\
HD\,94956           &  2.36&  1.07  &  45.3  &  424 &  192   & 3.0      & 0.8    & 2.8      & 1.0   \\
HD\,120170          & 23.90& 13.70  &  57.3  &   42 &   24   & 0.6      & 0.3    & 0.6      & 0.4   \\
HD\,132776          &  0.89&  0.99  & 111.2  & 1124 & 1250   & 4.6      & 3.0    & 4.0      & 4.9   \\
HD\,146116          &  1.16&  0.71  &  61.2  &  862 &  528   & 5.1      & 1.6    & 5.3      & 2.1   \\
HD\,188028          &  1.56&  0.78  &  50.0  &  641 &  321   & 4.3      & 1.1    & 4.3      & 1.5   \\
HD\,204046          &  1.39&  1.27  &  91.4  &  719 &  657   & 3.4      & 1.9    & 2.9      & 2.7   \\
HD\,207774          &  2.32&  1.30  &  56.0  &  431 &  242   & 2.7      & 0.9    & 2.4      & 1.2   \\
HR\,885             &  6.39&  0.35  &   5.5  &  157 &    9   & 3.6      & 0.1    & 3.6      & 0.1   \\
HR\,1023            &  2.68&  0.74  &  27.6  &  373 &  103   & 4.1      & 0.6    & 4.4      & 0.6   \\
HR\,1299            &  3.80&  0.39  &  10.3  &  263 &   27   & 3.6      & 0.3    & 3.5      & 0.3   \\
HR\,6757            &  4.65&  0.48  &  10.3  &  215 &   22   & 3.5      & 0.2    & 3.0      & 0.4   \\
HR\,6766            &  9.62&  0.26  &   2.7  &  104 &    3   & 3.6      & 0.1    & 3.6      & 0.1   \\
HR\,6791            &  7.96&  0.62  &   7.8  &  126 &   10   & 3.5      & 0.1    & 3.6      & 0.2   
\enddata
\end{deluxetable*}

\section{Abundances}
The focus of the abundance analysis is on those species that may provide clues to the reasons for
the weak G band, i.e., the significant underabundance of carbon relative to the slight
underabundance of carbon exhibited by giants which have passed through the first dredge-up.  If the
wk Gb stars' carbon underabundance is of nucleosynthetic origin, the H-burning CN-cycle is
presumably primarily responsible with the expectation that the carbon isotopic ratio $^{12}$C/$^{13}$C may
approach the cycle's equilibrium value of about 3 and the $^{14}$N will be overabundant but the sum
of the nuclides $^{12}$C + $^{13}$C + $^{14}$N will be unchanged from its value in the progenitors
of the wk Gb stars. A certain expectation for $^7$Li cannot be made but a difference between wk Gb
and normal giants is not unexpected. Although Li is readily destroyed by protons at the
temperatures required to run the CN-cycle, it may also be produced from existing $^3$He by the
Cameron-Fowler (1971) mechanism: $^3$He($^4$He)$^7$Be($e^-,\nu)^7$Li.  The atmospheric Li abundance
depends not only on the competition in the interior between production and destruction but also on
the transport of Li from the interior production (and destruction) site to the cooler exterior
layers where survival is assured.

The abundance analysis should also check the plausible assumption that the H-burning ON-cycles which
are effective at higher temperatures than the CN-cycle have not been a determining factor. With
optical spectra, this check is possible by determining the $^{16}$O abundance. Additional checks
from determinations of the $^{17}$O and $^{18}$O abundances may be possible when spectra of the
4.6$\mu$m CO ground state's fundamental vibration-rotation band are obtained. 

Abundances of other elements are not without interest. Demonstration that the abundances from Na to
Eu are normal would serve to restrict the possibility that diffusive processes have operated in a wk
Gb star's progenitor and might be a contributor to the severe carbon underabundance. Although one
might suppose the nucleosynthetic possibilities are remote, abundances of heavy elements should be
determined to check for unusual contributions from the neutron capture $s$-process (say, Y, La) and
$r$-process (say, Eu). An $s$-process enhancement, as in the case of Barium stars, may be possible
for a binary star in which a companion as an AGB star dumps material onto the present wk Gb star or
its progenitor but, however, the carbon underabundance is not readily explained by this idea.
Similarly, a massive companion exploding as a Type II supernovae would certainly leave its
nucleosynthetic imprint on the companion.

In the following sections, we describe the abundance analysis and its results. Atomic and molecular
data are described in the appropriate sections. At the outset, a comment needs to be made on the
molecular equilibrium which affects the partial pressures of C and O and very weakly of N. At the
temperatures of these giants, the only molecules of importance are CO and to a far lesser degree
N$_2$. Carbon monoxide with its well determined dissociation energy and partition function affects
the partial pressures of C and O in the outer layers of these giants.  Thus, these abundances must
be obtained iteratively. The derived C abundances are therefore reviewed after the determination of
the N and O abundance and vice versa until the changes in the abundances for each element are
smaller than 0.1\,dex (typically even lower than that). The N$_2$ molecule, also with a well
determined dissociation energy and partition function, has a very minor influence on the partial
pressure of N. The H$_2$ and H$_2$O molecules exert a totally negligible influence on the partial
pressure of H and O.

The abundance analyses are completed with both the spectroscopically and photometrically determined
atmospheric parameters. Although the abundances differ in detail, the key differences between wk Gb
stars and normal giants are unaffected and presented here for the first time for a complete sample
of known wk Gb stars accessible from the northern hemisphere. Interpreting the abundances is a major
challenge and not mitigated by the choice of the temperature scale.

For the synthesis of the different features described in the next sections we estimated the external
broadening parameters of the lines. We found that the effect of the combination of macroturbulence,
rotation, and instrumental profile can be well approximated by a Gaussian function. The fwhm of the
Gaussian is derived from different components. The fwhm of the instrumental profile can be
determined by measuring the fwhm of emission lines from ThAr spectra. The effect of macroturbulent
and rotational velocities is derived by measuring the fwhm of lines in different wavelength regions
of the echelle spectra. The determined fwhm for the Gaussian was then tested and adjusted by fitting
unblended lines in the vicinity of our lines of interest whenever possible.

The abundances derived in the following sections are summarized in Tables \ref{tab:cno} and
\ref{tab:abundances}. Table \ref{tab:cno} gives the abundances of Li, C, N, and O as well as the
$^{12}$C/$^{13}$C ratio. Elemental abundances are given for the spectroscopic and photometric
temperatures for each star. The C isotopic ratio is effectively independent of the temperature
scale. Table \ref{tab:abundances} gives abundances for Na to Eu for both the spectroscopic and
photometric temperatures.

\subsection{Carbon}\label{sect:carbon}
The C depletion of the typical wk Gb star is such that the carbon indicators in optical spectra most
often used in the analysis of normal GK giants are unavailable. For example,
\nocite{2007AJ....133.2464L}{Luck} \& {Heiter} (2007) in their large survey of nearby giants chose
the C\,{\sc i} 5380 and 6578 \AA\ lines and a feature of the C$_2$ Swan system at 5135\,\AA. In the
wk Gb stars, these lines are either absent or too weak to provide a useful abundance.  Although the
C\,{\sc i} triplet lines at 9078.28, 9088.51, and 9094.8\,\AA\ are present in our spectra, they are
blended strongly by telluric H$_2$O lines such that we do not consider them to be reliable abundance
indicators given our spectra. 

The CH G band, the A-X system,  near 4310\,\AA\ in normal giants is strong and saturated.
\nocite{1977ApJ...217..508L, 1981ApJ...248..228L}{Lambert} \& {Ries} (1977, 1981) adopted, as a
secondary C abundance indicator a CH feature at 4835\,\AA\ comprised of a blend of two 0-1 and a 1-2
A-X line with a strength of about 10\,m\AA\ in a typical and normal GK giant. This feature is
weakened below detectable limits in the wk Gb stars. By elimination, the potential C abundance
indicators in our optical spectra of wk Gb stars are reduced to CH lines comprising Fraunhofer's
original G band. The G band is not an ideal indicator because of the number of blending features
and uncertainties in establishing the continuum but fortunately the molecular data are well
determined.

The line list of molecular and blending atomic lines is taken from B. Plez (private communication)
for the considered region of 4300-4340\,\AA. The dissociation energy of CH is well determined: $D_0
= 3.465$ eV \nocite{Huber1979}(Huber \& Herzberg 1979). The $gf$-values of CH lines in Plez's list
are based  on band oscillator strengths derived from accurate experimentally determined radiative
lifetimes \nocite{larsson:4208}(Larsson \& Siegbahn 1986). \nocite{2005A&A...431..693A}{Asplund}
{et~al.} (2005)  provide a list of nine clean solar CH lines with wavelengths from 4218.72\,\AA\ and
4356.60\,\AA\ which yield a solar C abundance in satisfactory agreement with the abundance from
their primary C abundance indicators ([C\,{\sc i}], C\,{\sc i}, CH vibration-rotation, and C$_2$
Swan lines). Their adopted $gf$-values are in excellent agreement with those from Plez; the same
recipes appear to have been implemented. 

\begin{figure*}
\centering
\includegraphics[angle=-90]{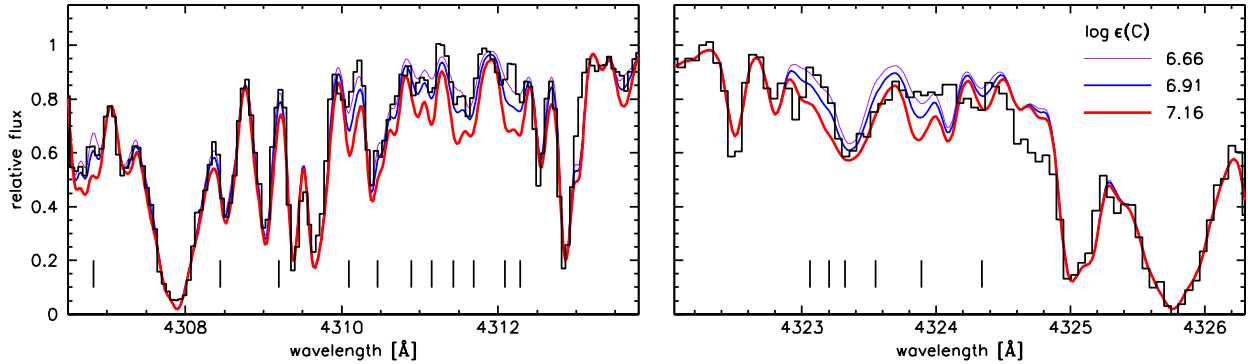}
\caption{A fragment of the CH G band in HD\,18636 at CH band heads (histogram). Overplotted are
synthetic spectra calculated with our spectroscopic temperatures. They show models with the best
fitting C abundance  of $\log \epsilon(\text{C})= 6.91$ and $\pm 0.25\,\text{dex}$. Key CH features
are marked with lines below the spectra. \label{fig:f3}}  
\end{figure*}

An example of the synthesis of a part of the G band for the wk Gb star HD 18636 is shown in Figure
\ref{fig:f3}.  No attempt has been made to fine tune the strengths of atomic lines across
the illustrated region; there are a few obvious wavelengths where the blending atomic lines are
either too strong or too weak. The best-fitting synthetic spectrum corresponds to a C abundance
[C/H] $\simeq -1.6$ when the O abundance is set at the value derived from the [O\,{\sc i}]
6300\,\AA\ line (see below). 

\subsection{Nitrogen}\label{sect:nitrogen}
The N abundance is obtained from lines of the CN red ($A^2\Pi - X^2\Sigma^+$) system and, in
particular, the 2-0 band near 8000\,\AA. The 1-0 band is seriously blended with telluric lines and
the 3-0 band is rather too weak to be useful.  The CN Violet system is too strong and blended to be
useful and often recorded with low S/N ratio on our spectra.  Of course, measurements of CN lines
demand  knowledge of the C (and O) abundances in order to obtain the N abundance.

A line list for the Red system was provided by B. Plez (2011, private communication). Wavelengths of
the useful lines are recomputed from energy levels given in
\nocite{Ram201082,2010ApJS..188..500R}{Ram}, {Wallace}, \&  {Bernath} (2010a); {Ram} {et~al.}
(2010b); all adopted lines have been measured off laboratory spectra. Plez's adopted $gf$-values are
based on theoretical quantum chemistry computations by \nocite{knowles:7334}Knowles {et~al.} (1988)
and \nocite{1988ApJ...332..531B}{Bauschlicher}, {Langhoff}, \&  {Taylor} (1988) in the absence of
definitive experimental determinations. \nocite{1998ApJ...508..387B}{Bakker} \& {Lambert} (1998)
review experimental and theoretical line strengths for the CN Red system and adopt the above quantum
chemistry calculations.  Their recommended  $gf$-values for the low excitation lines seen in
interstellar and circumstellar spectra are in excellent agreement with Plez's line list.

A long-standing issue for CN is the dissociation energy. We adopt $D_0= 7.72$ eV based on the
quantum chemistry calculations by \nocite{pradhan:3857}Pradhan, Partridge, \& Charles
W.~Bauschlicher (1994) aspects of  which were `calibrated' using calculations for C$_2$ for which an
accurate measure of its $D_0$ is available.  \nocite{pradhan:3857}Pradhan {et~al.} (1994) give
$D_0=7.72\pm0.04$ eV for CN and note that then recent experimental determinations by different
techniques give 7.77$\pm0.05$\,eV \nocite{1990A&A...232..270C}({Costes}, {Naulin}, \&  {Dorthe}
1990) and 7.738$\pm$0.02\,eV \nocite{doi:10.1021/j100180a079}(Huang, Barts, \&  Halpern 1992).  We
are unaware of more recent and more precise theoretical and experimental determinations of $D_0$ for
CN. 

In light of the history of uncertainties around the $gf$-values of the CN Red system and the
molecule's $D_0$, we investigated the solar CN 2-0 lines using the solar C and N abundances and
including the small correction for formation of CO molecules on the partial pressure of C. For this
purpose we interpolate from the \nocite{2004astro.ph..5087C}{Castelli} \& {Kurucz} (2004) model grid
a model with the solar parameters $\Teff = 5770\,\text{K}$, $\logg=4.40$, [Fe/H]$=0.02$, and $\vt =
0.8$\,km s$^{-1}$.  With the solar C abundance $\log\epsilon$(C) = 8.43 and $\log\epsilon$(O) = 8.69
from \nocite{2009ARA&A..47..481A}{Asplund} {et~al.} (2009) and the above choices for $gf$-values and
$D_0$, the solar 2-0 lines require a N abundance of 8.12 or a value 0.29\,dex higher than Asplund et
al.'s value from solar photospheric N\,{\sc i} lines. Asplund et al.  used their 3D models of the
solar photosphere and derived the C abundance provided from a selection of carbon lines, as noted in
the previous section.  Neither the CN Red nor the CN Violet system lines were used by Asplund et al.
An independent investigation of the solar CNO abundances with a different family of solar 3D model
atmospheres by \nocite{2011SoPh..268..255C}{Caffau} {et~al.} (2011) gives similar CNO abundances:
$\log\epsilon$(C) = 8.50 from a selection of C\,{\sc i} lines, $\log\epsilon$(N) = 7.86 from a
selection of N\,{\sc i} lines, and $\log\epsilon$(O) = 8.76 from a selection of O\,{\sc i} lines.

This apparent discrepancy between the solar N abundances from N\,{\sc i} lines and our result from
solar CN lines and published C abundances cannot be due entirely to systematic errors in the CN
molecular data, if the uncertainties have been correctly evaluated. The larger part of the
difference may reflect use of 3D solar models in the published analyses but we are using a 1D model
solar atmosphere. Comparisons published, for example, by \nocite{2005A&A...431..693A}{Asplund}
{et~al.} (2005)  for C$_2$ and CH and \nocite{2006A&A...456..675S}{Scott} {et~al.} (2006) for CO
suggest that use of  a 1D model may lead to a higher abundance  than when the 3D model is used and
that 1D-3D differences are larger for molecular lines than for atomic lines.  We are unaware of
analyses of CN Red system lines using 3D solar models.  Such analyses should be made. 

However, \nocite{2007AJ....133.2464L}{Luck} \& {Heiter} (2007) analysed the solar CN 6-2 and 7-3
lines near 6700 \AA\ assuming a similar $D_0$ (=7.65 eV), $gf$-values consistent Plez's list and
classical 1D model photospheres and reported a N abundance of about 8.17 with the precise value
dependent on the particular adopted solar model.  For their calculation, they adopted the C
abundance derived from solar C$_2$ lines and the same 1D model solar atmosphere and their C
abundance was similar to the 3D model-based results quoted above.  In the spirit of a differential
analysis of the wk Gb stars relative to the normal giants considered by Luck \& Heiter, we proceed
with our adopted CN data.

\begin{figure*}
\centering
\includegraphics[angle=-90]{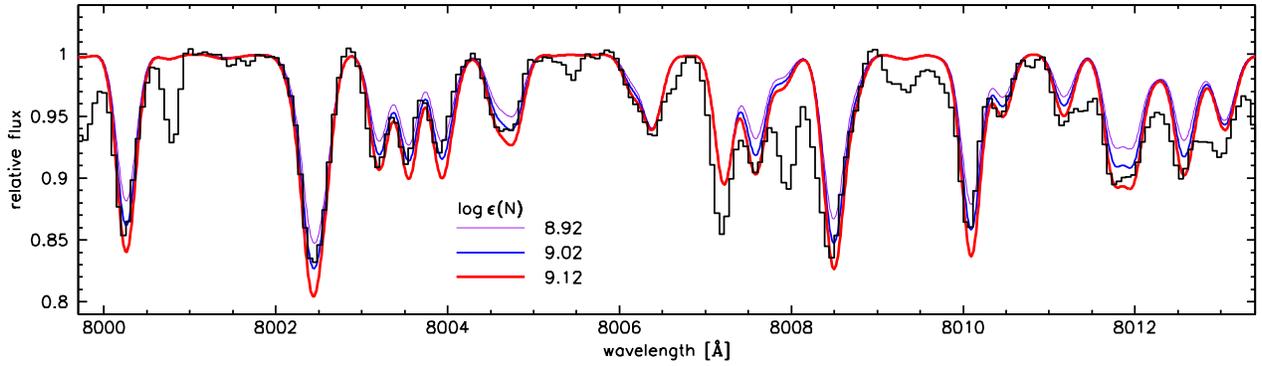}
\caption{The 2-0 band of the CN red system in HR\,6791. The C and O abundances are set at their best
values. The synthetic spectra are calculated for $ \log \epsilon (\text{N})= 9.02$ and $\pm
0.1\,\text{dex}$. \label{fig:f4}}  
\end{figure*}

Synthetic spectra of a portion of the CN 2-0 band are shown in Figure \ref{fig:f4} for
HR\,6791 with the C and O abundances set at their best values and spectra shown for three
choices of the N abundance. The $^{12}$C/$^{13}$C ratio is set at the value of 4 - see below.

\begin{figure}
\centering
\includegraphics[angle=-90]{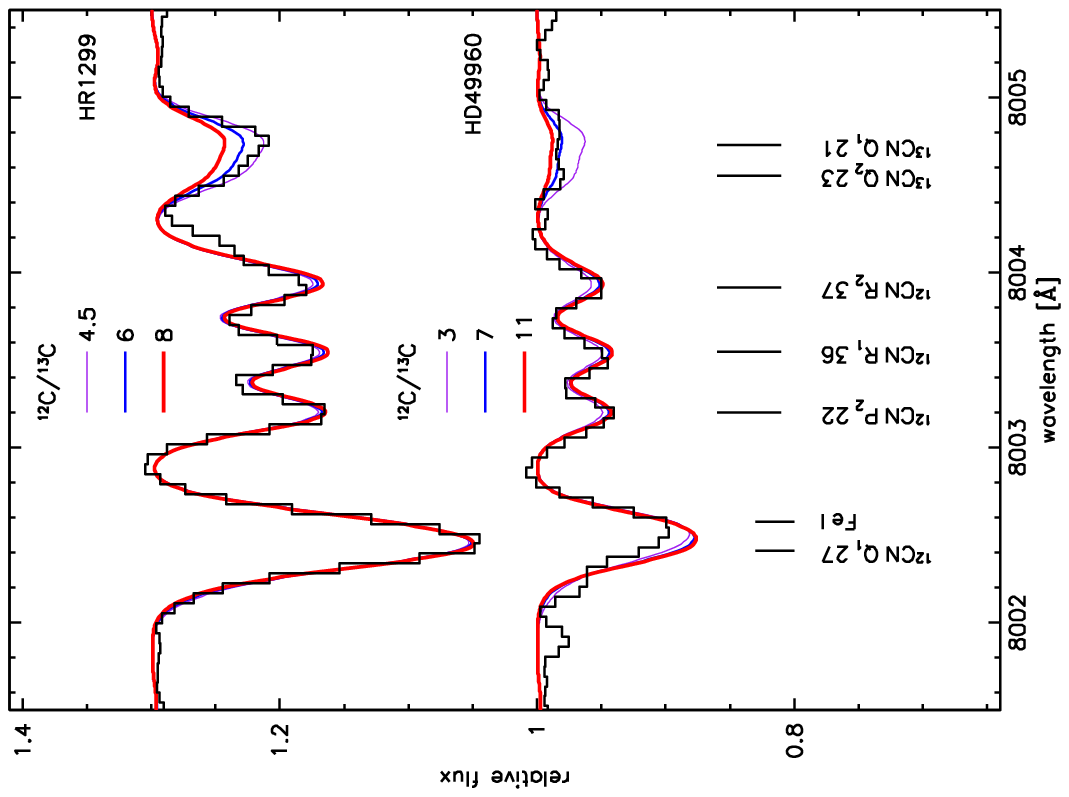}
\caption{Part of the 2-0 band of the CN red system in HD\,49960 and HR\,1299. For each object
synthetic spectra calculated with three different carbon isotope ratios are shown.
\label{fig:f5}}  
\end{figure}

The CN Red system is also a source of $^{13}$CN lines and, hence, of the important astrophysical
ratio $^{12}$C/$^{13}$C. Several $^{13}$CN lines appear unblended with stronger $^{12}$CN and/or
atomic lines and may be used for the isotopic abundance determination after, in some cases,
correction for overlying telluric H$_2$O lines.  Figure \ref{fig:f5} shows two examples of the
synthesis of a key $^{13}$CN feature.

\subsection{Oxygen}\label{sect:oxygen}
Two O abundance indicators are adopted: the [O\,{\sc i}] 6300 \AA\ line and the O\,{\sc i} triplet
at 7771.9, 7774.2 and 7775.4 \AA.

\begin{figure}
\centering
\includegraphics[angle=-90]{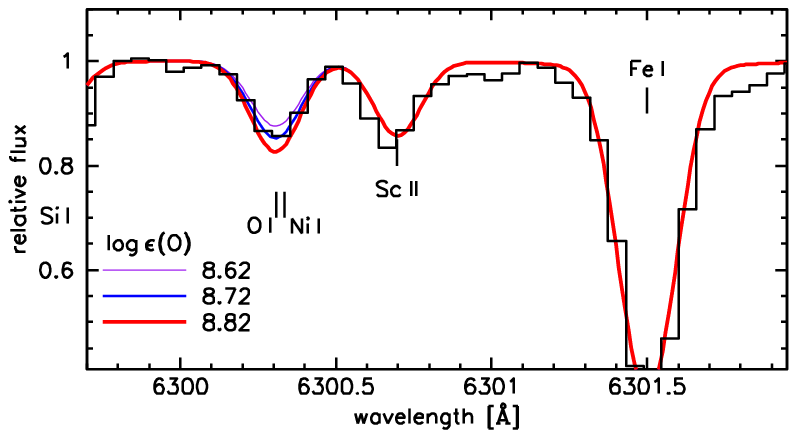}
\caption{The [O\,{\sc i}] 6300\,\AA\ line in HD\,18636. Overplotted are synthetic
spectra with $\log \epsilon(\text{O})=8.72$ and $\pm 0.1\,\text{dex}$. \label{fig:f6}}  
\end{figure}

The [O\,{\sc i}] line at 6300 \AA\ is blended with a Ni\,{\sc i} line \nocite{1978MNRAS.182..249L,
2001ApJ...556L..63A}({Lambert} 1978; {Allende Prieto}, {Lambert}, \&  {Asplund} 2001) for which we
took the $gf$-value from \nocite{2001ApJ...556L..63A}{Allende Prieto} {et~al.} (2001) and adopted
our derived Ni abundance (Table \ref{tab:abundances}). The $\log gf$-value of the [O\,{\sc i}] line
was set at -9.717 \nocite{2001ApJ...556L..63A}({Allende Prieto} {et~al.} 2001).  Figure \ref{fig:f6}
shows the line in HD\,18636 with synthetic spectra.

\begin{figure}
\centering
\includegraphics[angle=-90]{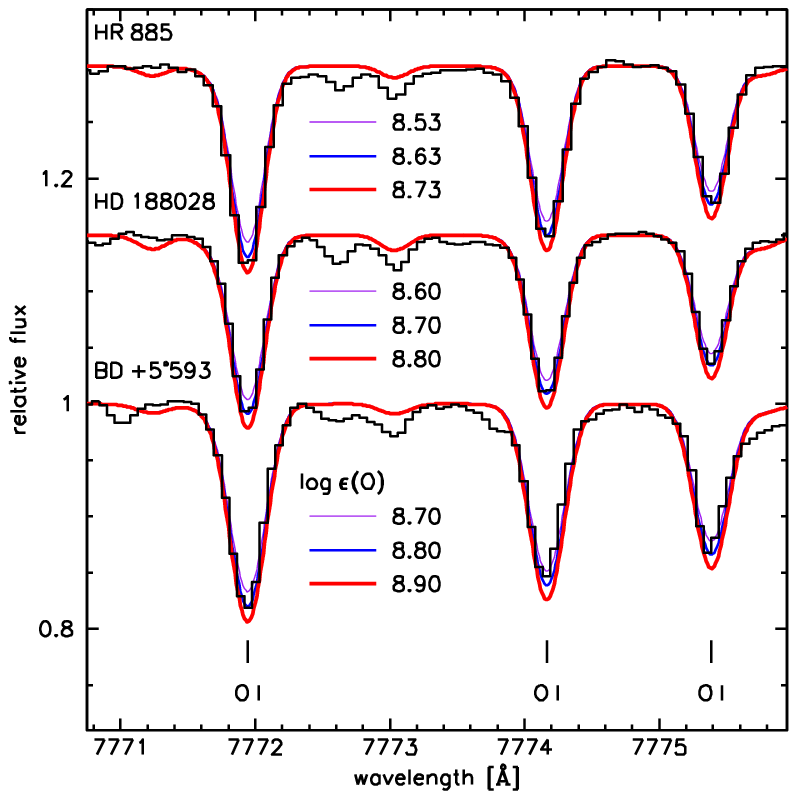}
\caption{The O\,{\sc i} triplet in three in wk Gb stars. Overplotted are synthetic
spectra with O abundances of $\log \epsilon(\text{O})= 8.63$ (HR\,885), 8.70
(HD\,188028), and 8.80 (BD\,+5$^{\circ}$593), and spectra with O abundances of $\pm
0.1\,\text{dex}$
for each star. The spectra of HD\,188028 and
HR\,885 are shifted upwards by 0.15 and 0.3. \label{fig:f7}}  
\end{figure}

The O\,{\sc i} triplet lines are sensitive to the adopted effective temperature, and affected by
non-LTE effects. The magnitude of the latter effects depends on the atmosphere parameters, the
oxygen abundance and differs for each line. For each star, we calculated non-LTE corrections for
each line individually with the IDL routine described in
\nocite{2007A&A...465..271R}{Ram{\'{\i}}rez}, {Allende Prieto}, \&  {Lambert} (2007).  These non-LTE
corrections agree with those from the grid calculated by \nocite{2009A&A...500.1221F}{Fabbian}
{et~al.} (2009) but cover a more suitable parameter range for our stars. Corrections are largest for
7771.9 \AA,  the strongest line of the triplet, with the mean correction over the sample being
$-0.36\pm0.07$ dex for the spectroscopic temperatures. For the weakest line (7775.4 \AA), the mean
correction is on average 0.06 dex smaller. Corrections for the photometric temperature scale are
larger by about 0.02 dex.  Figure \ref{fig:f7} shows observed and synthetic spectra of the oxygen
triplet for three wk Gb stars.

\begin{figure}
\centering
\includegraphics[angle=-90]{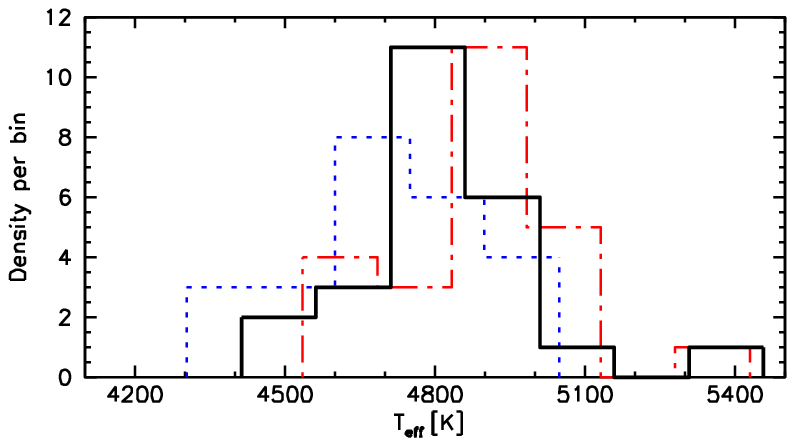}
\caption{Histogram of effective temperatures derived spectroscopically (dash-dotted line),
photometrically (dotted line), and by comparison of different oxygen lines (thick line). See text
for details. \label{fig:f8}}   
\end{figure}

The oxygen triplet shows a strong temperature sensitivity, while the forbidden 6300 \AA\ line's
temperature dependence is rather moderate. This makes it possible to perform an additional
temperature check by finding the effective temperature for each star at which the triplet and the
forbidden line give the same O abundance. Interestingly, the derived `oxygen' temperatures - see
Figure \ref{fig:f8} - suggest that the effective temperatures generally lie between our two
determinations.  

\nocite{2007AJ....133.2464L}{Luck} \& {Heiter} (2007) adopted the [O\,{\sc i}] 6300 \AA\ as their
sole indicator of the O abundance. The $gf$-value of the line was taken from
\nocite{2001ApJ...556L..63A}{Allende Prieto} {et~al.} (2001) with the $gf$-value of the blending
Ni\,{\sc i} line taken from the experimental determination by
\nocite{2003ApJ...584L.107J}{Johansson} {et~al.} (2003) and the Ni abundance set from the assumption
that [Ni/Fe] = 0. This recipe is very similar to ours such that we use Luck \& Heiter's O abundances
as the reference for normal GK giants subject to a small systematic difference arising from the use
of different model atmosphere grids. The O abundances quoted in Table \ref{tab:cno} and
discussed in the following sections are therefore the ones derived by the [O\,{\sc i}] 6300 \AA\
line.
\begin{deluxetable*}{lrrrrrrrrrr}
\tabletypesize{\footnotesize}
\tablecaption{Li and CNO abundances of the program stars.\label{tab:cno}}
\tablewidth{0pt}
\tablehead{
\colhead{Object}              &\colhead{[Fe/H]}            &\multicolumn{2}{c}{Li$_\text{NLTE}$}      &\multicolumn{2}{c}{C}       &
                               \colhead{$^{12}$C/$^{13}$C} &\multicolumn{2}{c}{N}       &\multicolumn{2}{c}{O}       \\ 
\colhead{}                    &\colhead{}                  &\colhead{sp.}&\colhead{ph.} &\colhead{sp.}&\colhead{ph.} &
                               \colhead{}                  &\colhead{sp.}&\colhead{ph.} &\colhead{sp.}&\colhead{ph.} 
}
\startdata
37\,Com             &   -0.53&    1.50&    1.16& 6.82& 6.78& $3.5\pm0.5$& 9.77& 10.13& 8.72& 8.67\\ 
BD\,+5$^{\circ}$593 &   -0.29&$<$ 0.19&$<$-0.04& 6.87& 6.77& $3\pm1$    & 8.85&  8.87& 8.80& 8.76\\
HD\,18636           &   -0.16&    1.93&    1.76& 6.91& 6.81& $8\pm2$    & 9.00&  9.02& 8.72& 8.69\\
HD\,28932           &   -0.39&    2.58&    2.27& 6.68& 6.56& $5\pm1$    & 8.87&  8.98& 8.61& 8.56\\
HD\,31869           &   -0.47&    1.06&    1.23& 6.77& 6.75& $4\pm1$    & 8.69&  8.68& 8.47& 8.46\\
HD\,40402           &   -0.14&    3.20&    3.06& 7.25& 7.20& $9\pm2$    & 8.84&  8.86& 8.81& 8.79\\
HD\,49960           &   -0.25&    0.82&    0.55& 6.85& 6.78& $7\pm1$    & 8.98&  9.08& 8.77& 8.73\\
HD\,67728           &   -0.43&$<$ 0.56&$<$ 0.28& 6.81& 6.70& $3.5\pm0.5$& 8.66&  8.67& 8.11& 8.08\\
HD\,78146           &   -0.06&$<$ 1.14&$<$ 0.80& 7.29& 7.21& $3\pm0.5$  & 8.83&  8.99& 8.77& 8.72\\
HD\,82595           &    0.06&    1.42&    1.22& 7.30& 7.21& $3\pm0.5$  & 8.86&  8.90& 8.73& 8.69\\
HD\,94956           &   -0.19&    1.21&    1.01& 6.93& 6.84& $5\pm1$    & 8.98&  9.02& 8.71& 8.68\\
HD\,120170          &   -0.51&    3.27&    3.01& 6.57& 6.48& $5\pm2$    & 8.73&  8.79& 8.54& 8.50\\
HD\,120171          &   -0.42&$<$-0.27&$<$-0.52& 7.53& 7.53& $5\pm1$    & 8.23&  8.22& 8.44& 8.44\\
HD\,132776          &   -0.07&    1.80&    1.41& 7.39& 7.33& $3.5\pm0.5$& 9.26&  9.57& 9.07& 9.01\\
HD\,146116          &   -0.39&$<$ 0.17&$<$-0.22& 6.73& 6.60& $3\pm0.5$  & 8.83&  9.00& 8.49& 8.44\\
HD\,188028          &   -0.18&$<$ 0.19&$<$-0.19& 6.70& 6.47& $4\pm1$    & 8.99&  9.09& 8.70& 8.58\\ 
HD\,204046          &   -0.05&$<$ 0.73&$<$ 0.36& 7.11& 6.97& $3.5\pm0.5$& 8.96&  9.09& 8.78& 8.72\\
HD\,207774          &   -0.33&$<$ 0.31&$<$ 0.02& 7.26& 7.16& $3.5\pm0.5$& 8.75&  8.79& 8.78& 8.74\\
HR\,885             &   -0.35&    1.46&    1.39& 6.48& 6.43& $5\pm1$    & 8.57&  8.56& 8.40& 8.39\\
HR\,1023            &   -0.22&    3.25&    2.91& 6.89& 6.39& $7\pm2.5$  & 9.37&  9.41& 8.61& 8.57\\
HR\,1299            &   -0.08&    2.85&    2.78& 7.25& 7.24& $4.5\pm0.5$& 9.12&  9.11& 8.62& 8.61\\
HR\,6757            &   -0.08&    2.36&    1.93& 7.18& 6.98& $3\pm0.5$  & 8.71&  8.88& 8.80& 8.74\\
HR\,6766            &   -0.18&    1.23&    1.08& 7.01& 6.93& $3.5\pm0.5$& 8.70&  8.71& 8.69& 8.67\\
HR\,6791            &    0.00&    1.99&    1.82& 7.47& 7.37& $4\pm0.5$  & 9.02&  9.00& 8.72& 8.67\\
\tableline
Sun                 &        & \multicolumn{2}{c}{3.26} & \multicolumn{2}{c}{8.43} & 89 & 
                               \multicolumn{2}{c}{7.83} & \multicolumn{2}{c}{8.69} 
\enddata
\tablecomments{Abundances of element A are given in $\log \epsilon(\text{A})$. Solar abundances are
taken from \nocite{2009ARA&A..47..481A}{Asplund} {et~al.} (2009). The errors for the different
abundances take into account the effect of uncertain parameters on the abundances and difficulties
in the synthesis of the lines, such as continuum placement and quality of the spectra. These errors
are added in quadrature. For N additionally the C error is added, since the only abundance indicator
- the CN red system - includes C. We estimate errors of $\sigma_\text{C}=0.25$,
$\sigma_\text{N}=0.3$, $\sigma_\text{O}=0.11$, and $\sigma_\text{Li}=0.25$.} 
\end{deluxetable*}

\paragraph{HD\,120171}
The star HD\,120171 shows a high C abundance compared to the rest of the program stars. The low
carbon deficiency is accompanied by an only moderate N overabundance and normal O abundance. Overall
this object does not show the striking characteristics of a wk Gb stars and might rather be
considered normal or intermediate between the normal giants and the wk Gb stars.

\subsection{Lithium}
\begin{figure}
\centering
\includegraphics[angle=-90]{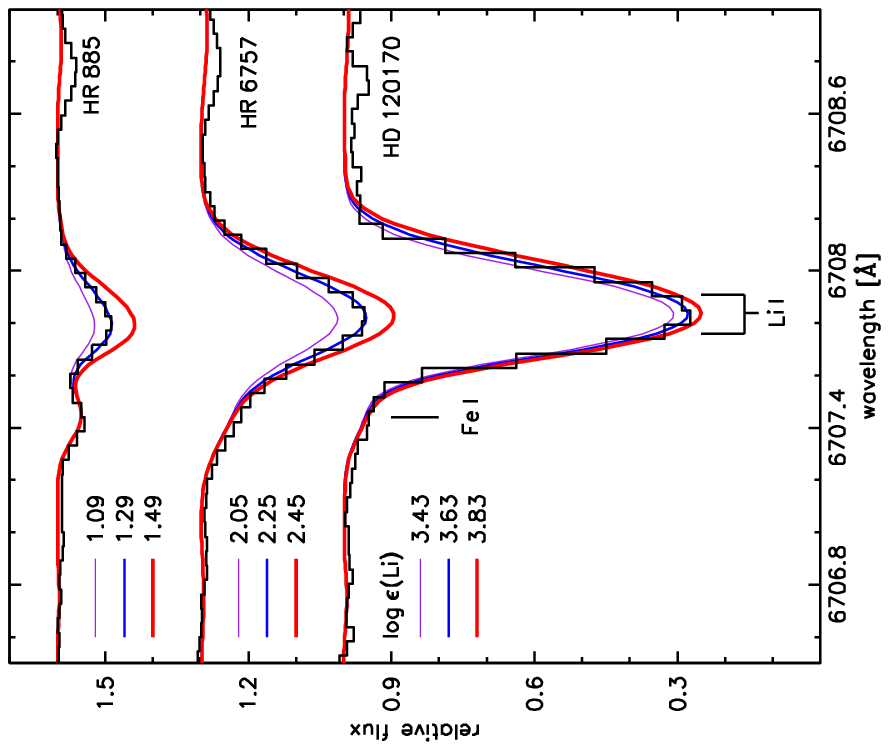}
\caption{Spectra of three wk Gb stars and overplotted synthetic spectra. The synthetic spectra
are calculated with Li abundances of $\log \epsilon(\text{Li})= 1.29$ (HR\,885), 2.25
(HR\,6757), and 3.63 (HD\,120170). Additional spectra with $\pm
0.2\,\text{dex}$ are also shown for each star. \label{fig:f9}}  
\end{figure}

Only one Li feature is prominent in our spectra, the 6708\,\AA\, Li\,{\sc i} resonance doublet. We
determine the Li abundance by synthesizing the feature and an Fe\,{\sc i} line at $6707.44\,\AA$.
The data used for the wavelength positions and \loggf values for $^7$Li and $^6$Li components of
this feature are taken from \nocite{1984ApJ...283..192L}{Lambert} \& {Sawyer} (1984). We include CN
blends in our synthesis and use the C and N abundances derived as described in sections
\ref{sect:carbon} and \ref{sect:nitrogen}. The precision of the Li abundance determination depends
on the strength of the lines. For weak lines a change of $\pm 0.05$ in the log already leads to
recognizable differences, while for strong lines sometimes a change of $\pm 0.2$ is necessary (see
Figure \ref{fig:f9} for an example of Li lines with different strength; the
Li\,{\sc i} line strength varies significantly even though all three stars have comparable
temperatures). Apart from the uncertainty
from the synthesis of the lines, the error for the Li abundances due to uncertainties of the model
atmosphere parameters is not likely to exceed 0.2\,dex. The biggest impact comes from an uncertain
temperature.  A change in \Teff of $\pm 200\,\text{K}$ leads to a change of 0.23\,dex in the Li
abundance. Both \logg and \vt have only a minor influence on the determined Li abundances. A
determination of the $^7$Li/\,$^6$Li isotope ratio is unpromising. For large Li abundances the
doublet blends with the close-by Fe\,{\sc i} line. The 6708\,\AA\, Li\,{\sc i} line can be affected
by non-LTE effects which can lead to positive or negative abundance corrections. The size of the
corrections depends on the stellar parameters and the lithium abundance itself
\nocite{2009A&A...503..541L}({Lind}, {Asplund}, \&  {Barklem} 2009). We calculated NLTE corrections
suitable for the parameters and Li abundances of the program stars by interpolating between the
tabulated values of \nocite{2009A&A...503..541L}{Lind} {et~al.} (2009).

The corrected Li abundances cover a wide range as can be seen in Table \ref{tab:cno}.    

\begin{figure}
\centering
\includegraphics[angle=-90]{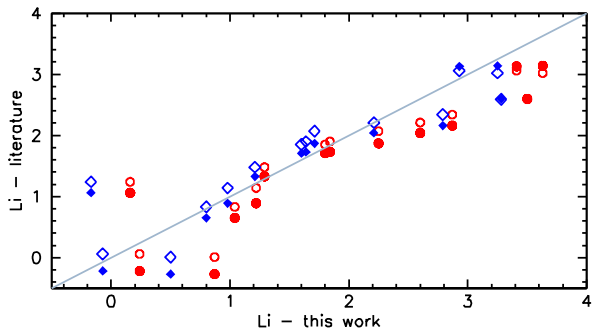}
\caption{Comparison of Li  LTE abundances with literature values. Circles represent the comparison
of our spectroscopic results with \protect\nocite{2012A&A...538A..68P}{Palacios} {et~al.} 2012
(filled) and the various results presented in \protect\nocite{1984ApJ...283..192L}{Lambert} \&
{Sawyer} 1984 (open). Analogously, squares indicate the comparison of our photometric results.
\label{fig:f10}}  
\end{figure}

Figure \ref{fig:f10} shows a comparison of our determined Li abundances with the values
\nocite{2012A&A...538A..68P}{Palacios} {et~al.} (2012) and \nocite{1984ApJ...283..192L}{Lambert} \&
{Sawyer} (1984). Differences between the two literature sources are naturally small, since both are
using the same equivalent widths of the Li lines. Our newly determined Li abundances in general
agree well with the literature data. A larger scatter can be recognized at lower Li abundances,
mainly due to the more difficult detection and analysis of these Li lines, which only allows the
determination of upper limits.

\subsection{Other elements}\label{sect:otherelements}
Abundances for Na, Mg, Al, Si, Ca, Ti, Cr, Fe, Ni, Y, and Eu were determined. Most of these elements
appear in the spectra with a sufficient number of unblended lines. Their abundances can be derived
by force-fitting the abundances of the lines to match measured equivalent widths as described in
Sect.  \ref{sect:parameters}.  The atomic data for all lines used in the analysis is taken from
\nocite{2007AJ....133.2464L}{Luck} \& {Heiter} (2007). We attempted to perform our analysis as
similar as possible to Luck \& Heiter's and to utilize as many of their lines as possible. Not all
of the lines were present in our objects but no additional lines from other sources were added. We
determine solar abundances for each of the lines used and derive relative abundances for the
elements in the wk Gb stars on a per-line basis. 
\begin{figure*}
\centering
\includegraphics[angle=-90]{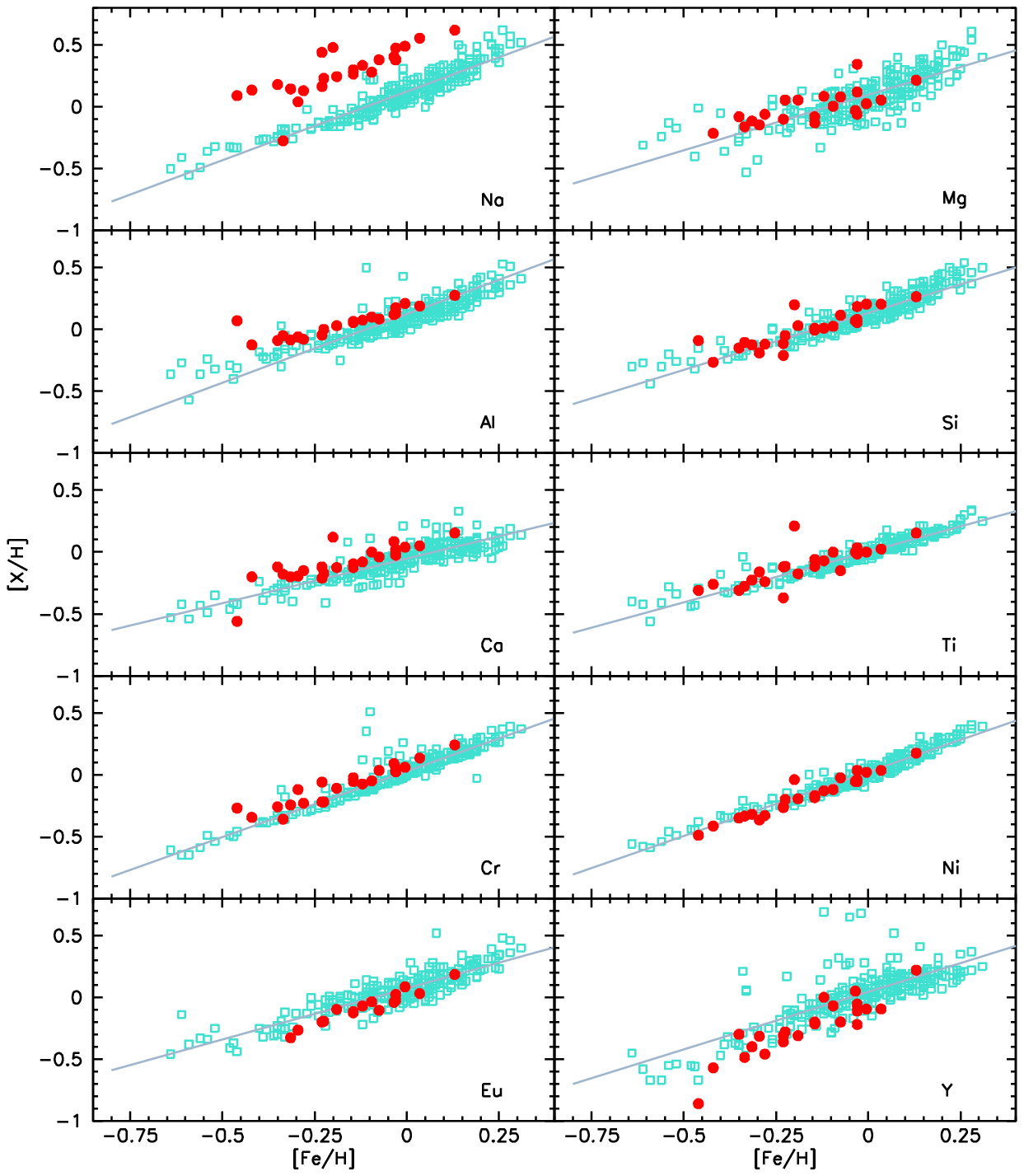}
\caption{Trends of element abundances vs. metallicity in the wk Gb stars for spectroscopic
temperatures (filled circles), compared to giants from \protect\nocite{2007AJ....133.2464L}{Luck} \&
{Heiter} 2007 (open squares).  The solid lines indicate the slopes of the trends from the
spectroscopic analysis of \protect\nocite{2007AJ....133.2464L}{Luck} \& {Heiter} (2007).
\label{fig:f11}}  
\end{figure*}

\paragraph{Na}
We measured equivalent widths of Na\,{\sc i} lines at 6154.225 and 6160.747\,\AA\ in our spectra. 

The Na abundance is enhanced in the wk Gb stars with $[\text{Na/Fe}]=0.46\pm0.12$ \footnote{Relative
to the solar abundances derived with the line list described in section \ref{sect:otherelements}.}
for the spectroscopic temperatures and $[\text{Na/Fe}]=0.37\pm0.13$ for the photometric
temperatures. We determined a [Na/Ca] ratio of $0.34\pm0.15$ (sp.) and $0.37\pm0.20$ (ph.).
\nocite{1994ApJ...435..797D}{Drake} \& {Lambert} (1994) determined $[\text{Na/Ca}]=0.16$ for a
smaller sample of the wk Gb stars. A direct comparison with their result, however, is not easily
possible since their abundance data was derived relative to Pollux. For the objects that both
samples have in common the EW that we measured for the Na lines are in excellent agreement with the
ones measured by \nocite{1994ApJ...435..797D}{Drake} \& {Lambert} (1994). The mean differences are
$-1.9\pm 5.95\,\text{m\AA}$ for the 6154\,\AA\ line and $1.57\pm 10.78\,\text{m\AA}$ for the
6160\,\AA\ line. The differences in the Na abundances between our and Drake \& Lambert's analysis
then depend on the differences in model parameters and the reference to Pollux.

The comparison with the abundances from \nocite{2007AJ....133.2464L}{Luck} \& {Heiter} (2007) shows
a significant enhancement of Na in the wk Gb stars compared to the normal giants by about 0.3\,dex
(see Figure \ref{fig:f11}).  Note also that the only star that shows a Na comparable with the data
from \nocite{2007AJ....133.2464L}{Luck} \& {Heiter} (2007) is HD\,120171, whose C and N abundance
are closest to those of normal giants. A significant enrichment of Na in the rest of the wk Gb stars
could be a result of a large amount of $^{22}$Ne converted to $^{23}$Na.

\paragraph{Mg, Al, Ca, Si, Ti, Cr, Ni, Eu, and Y}
We determined the abundances of a group of additional elements, including the $\alpha$-elements Mg,
Ca, Si, and Ti, the $r$-process element Eu and the $s$-process element Y, using again the line data
from \nocite{2007AJ....133.2464L}{Luck} \& {Heiter} (2007). In Figure \ref{fig:f11} we compare these
abundances with the abundances from \nocite{2007AJ....133.2464L}{Luck} \& {Heiter} (2007). Contrary
to Na (see above) the abundances of the elements from Mg to Y in the wk Gb stars are consistent with
the ones for the normal giants from \nocite{2007AJ....133.2464L}{Luck} \& {Heiter} (2007). 

The abundance trends of the elements found for the wk Gb stars (except Na) supports the conclusion
that wk Gb stars are normal in all other respects than their CNO-cycle element and Li abundances. 

\section{Discussion}\label{sect:discussion}
\subsection{Preliminary remarks}\label{sect:prelremarks}
Compositions of wk Gb stars will be compared and contrasted with results for nearby normal GK giants
presented by \nocite{2007AJ....133.2464L}{Luck} \& {Heiter} (2007) whose abundance analysis closely
resembles ours, particularly with respect to selection of lines and atomic/molecular data. It seems
apparent, as noted above, that the wk Gb stars have experienced substantial exposure to the
H-burning CN-cycle. After a few preliminaries, it is this aspect which we discuss in the next
section. 

\begin{figure}
\centering
\includegraphics[angle=-90]{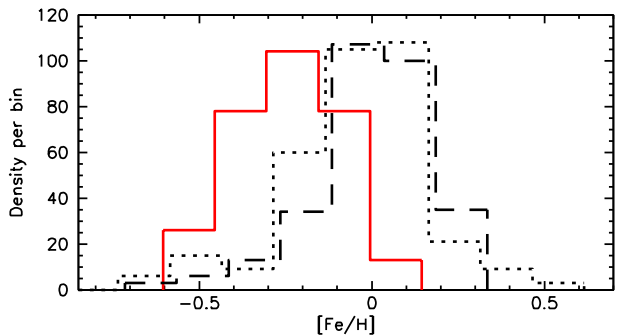}
\caption{Metallicity distribution in the wk Gb stars (solid line) compared to the [Fe/H]
distributions from the giants sample in \protect\nocite{2007AJ....133.2464L}{Luck} \& {Heiter}
(2007) (dashed line) and the dwarf sample from \protect\nocite{2006AJ....131.3069L}{Luck} \&
{Heiter} (2006) (dotted line). The distributions for the wk Gb stars and the dwarf data were
scaled to the number of giants from \protect\nocite{2007AJ....133.2464L}{Luck} \& {Heiter} (2007).
\label{fig:f12}}  
\end{figure}

One of the preliminary remarks concerns the [Fe/H] distributions of our and Luck \& Heiter's sample.
These are compared in Figure \ref{fig:f12} and show that the wk Gb stars have a higher proportion of
low [Fe/H] stars (say, [Fe/H] $< -0.2$) than do the nearby giants which Luck \& Heiter show have the
same distribution as the nearby dwarfs \nocite{2006AJ....131.3069L}({Luck} \& {Heiter} 2006) with
both giant and dwarf distributions having mean values [Fe/H] $\simeq 0.00$. In contrast, the wk Gb
sample has a mean value $[\text{Fe/H}] = -0.23$.  Unless the search for wk Gb stars was concentrated
more heavily on metal-poor giants, this result implies that production of wk Gb stars is more
readily accomplished among metal-poor stars. 

A second preliminary remark concerns the stellar masses. The normal giants run from about 1\,\Msun
to about 3\,\Msun with the majority of stars falling between 1 and 2\,\Msun
\nocite{2007AJ....133.2464L}({Luck} \& {Heiter} 2007). In contrast, the
trigonometrical parallaxes and effective temperatures  place the masses  of our wk Gb sample (see
above)  between 2.5\,\Msun and about 5\,\Msun, i.e., our stars are what are usually referred to as
intermediate mass stars and the nearby stars are low mass stars. This mass difference translates to
differences in the evolution of the stars, e.g., low mass but not intermediate mass stars experience
a He-core flash. But a more significant difference may be the differing distributions of rotational
velocity. As \nocite{1967ApJ...150..551K}{Kraft} (1967) showed, the mean rotational velocity
declines from high values on the upper upper main sequence stars ($M > 1.25\,\Msun$) to low values
on the lower main sequence. Thus, the wk Gb stars may have left the main sequence as rapid rotators
but the majority of Luck \& Heiter's sample of giants departed the main sequence as slow rotators.
If rapid rotation of an intermediate mass star is responsible for the creation of a wk Gb star, one
may wonder why such a small minority of intermediate mass main sequence stars evolve to wk G stars.   

The third preliminary remark concerns the signatures of the H-burning, by the CN-cycle and possibly
the ON-cycles. All normal giant stars are expected to have experienced the first dredge-up which
brought mildly CN-cycled material into their atmosphere. Low mass stars may have also experienced
additional mixing at the luminosity bump on the red giant branch or, perhaps,  at the He-core flash.
Luck \& Heiter's sample includes stars on the red giant branch (pre-He-core flash) and stars at the
red clump (post-He-core flash). Intermediate mass giants experience  first and second dredge-up but
not the He-core flash. 

Signatures of the addition of mildly CN-cycled material to an atmosphere include a slight reduction
of $^{12}$C, increase of $^{13}$C, increase of $^{14}$N with the constraint that the sum
$^{12}$C+$^{13}$C+$^{14}$N is conserved. In the first dredge-up for low mass stars, the $^{16}$O
abundance is not changed measurably. Changes of the $^{17}$O and $^{18}$O abundance are expected.
It is unfortunate that these two heavier O isotopes are not accessible from optical spectra. In the
absence of production by the Cameron-Fowler mechanism, the atmosphere's Li abundance will be lowered
severely. 

These predictions are confirmed by Luck \& Heiter's sample, as they noted. In particular, comparison
of results for nearby dwarfs and giants showed C in the giants was lowered by about 0.2 dex, N was
increased in giants over levels seen in dwarfs such that the C+N sum was conserved, and O abundances
among dwarfs and giants were similar. (The spectral coverage did not allow  Luck \& Heiter to
determine the carbon isotopic ratio but results in the literature show that the ratio is lowered in
giants in qualitative, if not quantitative, agreement with expectation.)

\begin{figure*}
\centering
\includegraphics[angle=-90]{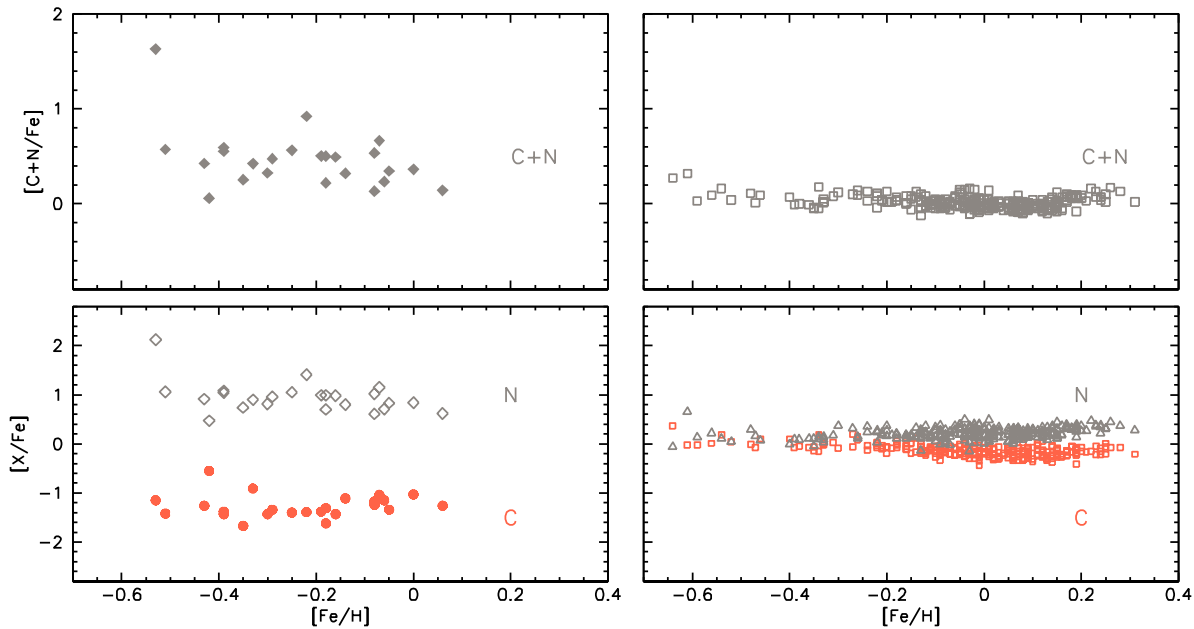}
\caption{Lower left panel: Spectroscopically derived abundances of C (filled circles) and N (open
squares) for the wk Gb stars. Lower right panel C (open squares) and N (open triangles) data from
\protect\nocite{2007AJ....133.2464L}{Luck} \& {Heiter} (2007). Upper left panel: Sum of C and N
abundances in the wk Gb stars. Upper right panel: The same for the giants from
\protect\nocite{2007AJ....133.2464L}{Luck} \& {Heiter} (2007). For this plot we use the solar
abundances for C and N as derived by \protect\nocite{2007AJ....133.2464L}{Luck} \& {Heiter} (2007)
with the newest MARCS models: $\log\,\epsilon (\text{C})=8.50$, $\log\,\epsilon (\text{N})=8.18$.
\label{fig:f13}}  
\end{figure*}

Carbon and nitrogen abundances for wk Gb and normal giants are compared in Figure \ref{fig:f13} with
the left-hand panel presenting our results and the right-hand panel showing results for nearby
normal giants from \nocite{2007AJ....133.2464L}{Luck} \& {Heiter} (2007). 

A direct  way to examine the assertion that the wk Gb stars are unusually rich in CN-cycled
material is to plot in Figure \ref{fig:f13} the C and N abundances  versus [Fe/H] for the wk Gb
stars (left-hand lower panel) and normal nearby giants (right-hand lower panel). The
contrast between the two samples is striking: the wk Gb stars are about a factor of 20 underabundant
in C relative to normal giants. Additionally the sum of C and N (here C is the sum of $^{12}$C and
$^{13}$C) is about 0.3 dex greater on average for wk Gb stars than for the nearby giants over a
common range in [Fe/H]. A possible explanation for this difference is that, in addition to excessive
CN-cycling, some $^{16}$O has been processed to $^{14}$N through partial ON-cycling. The fact that
the $^{12}$C/$^{13}$C ratios are very low is also evidence for CN-cycling.

In terms of the C and N abundances, there appears a clean distinction between normal and wk Gb
giants. This, however, may be an artefact because the discovery of wk Gb stars has come almost
entirely from visual  inspection of low dispersion spectra. The star with a seemingly intermediate C
abundance is HD\,120171 (see section \ref{sect:oxygen}).

\begin{figure}
\centering
\includegraphics[angle=-90]{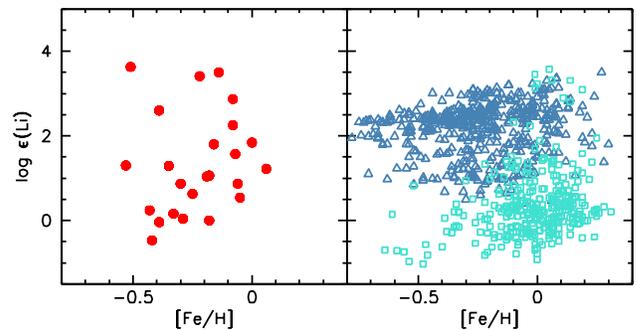}
\caption{Spectroscopic LTE Li abundances of wk Gb stars (left panel) compared to giants from
\protect\nocite{2007AJ....133.2464L}{Luck} \& {Heiter} (2007) (squares, right panel) and dwarf data
from \protect\nocite{2004MNRAS.349..757L}{Lambert} \& {Reddy} (2004) (triangles, right panel).} 
\label{fig:f14}  
\end{figure}

Lithium in normal giants is generally severely depleted, as predicted because the giant's convective
envelope dilutes the lithium that has survived in the main sequence star, the giant's progenitor.
This dilution is illustrated in the right-hand panel of Figure \ref{fig:f14} where Li abundances for
Luck \& Heiter's giants are contrasted with Li abundances for dwarfs from
\nocite{2004MNRAS.349..757L}{Lambert} \& {Reddy} (2004). The non-LTE Li abundances for wk Gb stars
(Table \ref{tab:cno}) span the range from the interstellar abundance ($\log
\epsilon(\text{Li})\simeq 3.3)$, the star's likely initial Li abundance, to severely depleted
abundances (see also left-hand panel of Figure \ref{fig:f14} for a comparison of wk Gb LTE
abundances with normal giants and dwarfs). The most direct interpretations of this 1000-fold minimum
spread in Li abundances is that Li dilution by the giant's convective envelope is in some cases
offset by internal synthesis of Li to an extent that varies greatly from star-to-star despite
comparable degrees of contamination by CN-cycle products. 

\subsection{Clues from nucleosynthesis}
To unravel the links between the composition of wk Gb stars and their origins calls for -- first --
dissection of the nuclear processes to which their atmospheres has been exposed and -- second  --
identification of the mechanism responsible for transport of the products of these nuclear processes
from their site of operation to the stellar atmosphere. 

A high N/C ratio and a  low $^{12}$C/$^{13}$C ratio are the principal signatures that H-burning
CN-cycle products now contaminate a wk Gb star's atmosphere. In parallel with the CN-cycle, the
slower ON-cycles and the NeNa- and MgAl-chains may run with observable consequences.  Finally,
lithium production by the \nocite{1971ApJ...164..111C}{Cameron} \& {Fowler} (1971) mechanism
converts available $^{3}$He via $^7$Be to $^7$Li. Since the atmospheric Li abundance depends greatly
on the efficiency of transfer of the $^7$Be and $^7$Li to low temperatures in order to avoid
destruction by protons, prediction of a wk Gb star's Li abundance is fraught with uncertainty. 

Given that the $^3$He reservoir is potentially capable of synthesizing Li to levels far greater
than are observed in wk Gb stars, it is intriguing that the maximum observed Li abundance is
coincident with the interstellar Li abundance - see \nocite{1984ApJ...283..192L}{Lambert} \&
{Sawyer} (1984) for a speculative interpretation of this fact. Lithium in the rare examples of
Li-rich (normal) giants can exceed the interstellar abundance by a factor of several
(\nocite{2000A&A...359..563C}{Charbonnel} \& {Balachandran} 2000;
\nocite{2011ApJ...730L..12K}{Kumar}, {Reddy}, \& {Lambert} 2011). Thus, little weight should be
given the above coincidence. 

In contrast to the case of lithium, the relative abundances from the H-burning cycles and
chains are predictable with fair certainty from nucleosynthetic arguments, even if aspects of the
stellar evolutionary context remain murky.

Consumption of protons by the (slow) CN-cycle converts four protons to an $\alpha$-particle and,
after a few cycles, establishes equilibrium abundances of the participating nuclides: $^{12}$C,
$^{13}$C, $^{14}$N and $^{15}$N with broadly $^{12}$C/$^{13}$C $\sim 3$, $^{14}$N/$^{12}$C $\sim$
100, and $^{14}$N/$^{15}$N $>$ 20,000 at the temperatures of efficient H-burning. Before presenting
a semi-quantitative interpretation of the wk Gb star CNO abundances, one may note that the
abundances (Table 4) mirror the above estimates: the N/C ratio is about 100, and the carbon isotopic
ratio is 3-4 for many stars. 

Two boundary prerequisites for the preeminence of the CN-cycle must be recognized. First, the
CN-cycle is in competition for protons with the $pp$-chains. Second, the ON-cycles process protons
more slowly than the CN-cycle and, thus, there is a possibility that the CN-cycle and the $pp$-chain
will complete the conversion of H to He before the ON-cycles will have had time to imprint their
effect on the CNO abundances. 

Our evaluation of these two conditions is made for a typical wk Gb star with [Fe/H] $\sim -0.3$ and
initial CNO abundances (i.e., in the progenitor dwarf) of approximately C = 8.2, N = 7.5 and O = 8.6
\footnote{Here element $\text{A} \equiv \log \epsilon (\text{A})$}. Nuclear
reaction rates are taken from the NACRE compilation \nocite{1999NuPhA.656....3A}({Angulo} {et~al.}
1999); recent updates \nocite{2011RvMP...83..195A}({Adelberger} {et~al.} 2011) are unimportant for
what follows. 

The  relative efficiencies of the $pp$-chains and the CN-cycle in consuming protons are set by the
slowest rate for each process: $p(p,e^+\nu_e)d$ for the $pp$-chain and $^{14}$N($p,\gamma)^{15}$O
for the CN-cycle. As is well known, the $pp$-chains are the principal consumer of protons at `low'
temperatures and the CN-cycle at `high' temperatures. For N = 7.8, the consumption rates are equal
at about 20 million degrees. Thanks to the Coulomb barrier  between the $^{14}$N and the proton, the
$pp$-chain rapidly outperforms the CN-cycle as the temperature is lowered below 20 million degrees;
for example, the $pp$-chain is favored by a factor of $10^5$ at 10 million degrees. Thus, the
CN-cycled material present in a wk Gb star was processed at temperatures of 20 million degrees or
hotter. 

The ON-cycles consume protons  more slowly than the CN-cycle and are unlikely to achieve their
equilibrium abundances.  Flow through the ON-cycles is controlled by the ratio of the reaction rates
$^{15}$N$(p,\alpha)^{12}$C which closes the CN-cycle and $^{15}$N$(p,\gamma)^{16}$O which opens the
ON-cycles. This ratio reflecting the relative strengths of the strong to the electromagnetic
interaction favors closure of the CN-cycle by a factor of about 1000.  However, in advance of the
ON-cycles reaching equilibrium, some $^{16}$O is converted to $^{14}$N via $^{17}$O.  The
temperature-dependent CN-cycle equilibrium ratio for N/C may be increased by N from O consumption as
$^{16}$O is processed to $^{14}$N ahead of equilibrium within the ON-cycles.   

Given this contrast between the times for completion of the CN- and ON-cycles, one may note two
reasons for expecting the CN-cycle products in a wk Gb stellar atmosphere not to be accompanied by
ON-cycle equilibrium products which are $^{16}$O-poor and $^{14}$N-rich.  First, the episode of
H-burning may  not be active for the length of time required to run the ON-cycle yet able to run the
CN-cycle to equilibrium. Second, protons may be totally consumed by the CN-cycle before the ON-cycle
achieves equilibrium. For example, for our typical wk Gb star with C+N  = 8.3 and H/C = 3.7 and four
protons consumed per CN-cycle, 1200 (=$10^{3.7} - 10^{0.6}$) cycles exhaust the proton supply which
would indicate that the ON-cycle being a factor of 1000 slower than the CN-cycle will be largely
idle. (At much lower metallicities when the number of C+N catalysts is lower, the CN-cycle must run
more cycles to exhaust the protons and, then, O-poor (and C-poor) but very N-rich stars may be
anticipated.) A corollary deserves comment. If the ON-cycle has operated in the material mixed into
the atmosphere, the atmosphere will become H-poor and He-rich but the wk Gb stars are obviously not
extremely He-rich. An accurate determination of their He/H ratio is an elusive goal. 

\begin{figure}
\centering
\includegraphics[angle=-90]{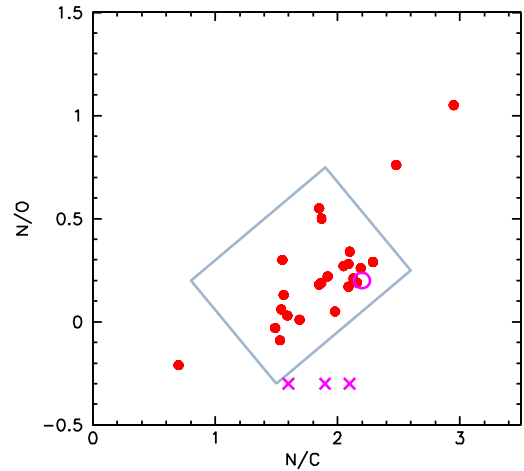}
\caption{N/O vs. N/C for the wk Gb stars (filled circles). Predicted N/C and N/O ratios are marked
with crosses. The N/O ratio derived by partial conversion of $^{16}$O to $^{14}$N is marked
with an open circle. See text for details.
\label{fig:f15}}  
\end{figure}

Now consider the typical wk Gb star with initial abundances C = 8.2, N = 7.5, and O= 8.6 or N/C =
-0.7 and N/O = -1.1. Suppose the atmosphere is mixed with a large amount of CN-cycled material, the
atmospheric composition tends to equilibrium abundances of the cycle. Then, the N/C becomes 2.1, 1.9
and 1.6 for temperatures T$_6$ = 20, 30 and 50, respectively, with  N/O = $-0.3$ for these
temperatures.\footnote{T$_6$ is $10^{6}$T K.} In Figure \ref{fig:f15}, we show N/O
versus N/C. With the exceptions of 37\,Com and HR\,1023 having high N/C and N/O, and HD\,120171 with
its high C and low N abundances, the wk Gb stars fall in a slanted rectangle centered on about N/O
$\simeq 0.3$ and N/C $\simeq 2$. The predicted N/Cs cross the slanted rectangle but the N/O ratio
does not: N/O = $-0.3$ is predicted but the stars span the range 0.0 to $+$0.6.  One way to raise
the N/O ratio is to assume that the ON-cycles have partially converted $^{16}$O to $^{14}$N in the
run up to the never attained condition of ON-cycle equilibrium. Suppose that at T$_6$ = 30 the O
abundance was depleted from 8.6 to 8.4, then  N/O = 0.2 and N/C = 2.2  which combination falls
within the rectangle (open circle in Figure \ref{fig:f15}). 

\begin{figure}
\includegraphics[angle=-90]{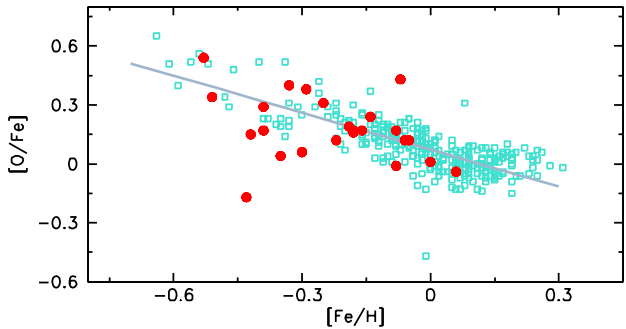}
\caption{Spectroscopic abundance of oxygen vs. iron abundance in the wk Gb stars (filled circles)
compared to the oxygen abundance of normal giants (open squares,
\protect\nocite{2007AJ....133.2464L}{Luck} \& {Heiter} 2007). The grey line is a linear fit to the
normal giants oxygen abundances. \label{fig:f16}}  
\end{figure}

\begin{figure}
\centering
\includegraphics[angle=-90]{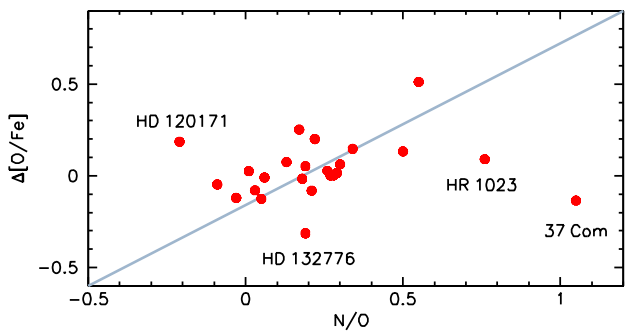}
\caption{$\Delta$[O/Fe] versus N/O for the wk Gb stars. The grey line indicates an assumed linear
correlation and is not a fit to the data points. Objects with peculiar values are labeled and
explained in the text.
\label{fig:f17}}  
\end{figure}

There is evidence for a star-to-star variation in O depletion. In Figure \ref{fig:f16}, we
compare [O/Fe] for wk Gb stars and normal giants with a line corresponding to a fit to the latter
results. If O depletion is the explanation for the N/O discrepancy, there will be a correlation
between the O deficiency as shown in  Figure \ref{fig:f16} by the amount that [O/Fe] falls
below the line defined by the normal giants (say, $\Delta$[O/Fe]) and the N/O ratio. Figure
\ref{fig:f17} shows the $\Delta$[O/Fe] versus N/O relation with the majority of points
exhibiting a loose correlation of the expected sign. There are four outliers: 37\,Com, HR\,1023,
HD\,132776 with high N abundances and HD\,120171 with a low O abundance but not the anticipated N
overabundance. HD\,67728 with a large negative $\Delta$[O/Fe] ($-0.6$) may also be an outlier.

\begin{figure}
\centering
\includegraphics[angle=-90]{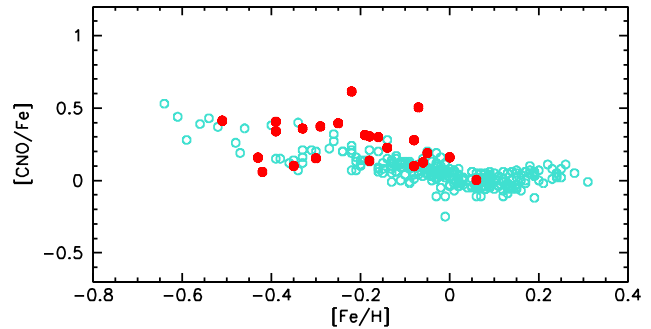}
\caption{Sum of C, N, and O abundances in the weak Gb stars (filled circles) compared to the giants
from \protect\nocite{2007AJ....133.2464L}{Luck} \& {Heiter} (2007) (open circles). For this plot we
use the solar abundances for C, N, and O as derived by \protect\nocite{2007AJ....133.2464L}{Luck} \&
{Heiter} (2007) with the newest MARCS models: $\log\,\epsilon (\text{C})=8.50$, $\log\,\epsilon
(\text{N})=8.18$, and $\log\,\epsilon (\text{O})=8.81$.
\label{fig:f18}}  
\end{figure}

If the CN-cycle supplemented by partial operation of the ON-cycles has occurred, a sum rule
operates, i.e.,  the sum of initial C, N, and O nuclei is conserved. This rule may be checked using
the nearby giants to define the initial sum. Figure \ref{fig:f18} shows the CNO sum versus [Fe/H]
for wk Gb stars and the nearby giants. Apart from two outliers among the wk Gb stars (37\,Com and
HR\,1023 \footnote{Both stars are rapid rotators. \nocite{2002AJ....123.2703D}{Drake} {et~al.}
(2002) give $v \sin i=11.0$ for 37\,Com and \nocite{1999A&AS..139..433D}{de Medeiros} \& {Mayor}
(1999) give $v \sin i=22.7$ for HR\,1023.}), the sum for the wk Gb stars may be somewhat
systematically greater than for the nearby giants; the upper envelope for the wk Gb stars  is
approximately 0.1\,dex greater than for nearby giants of a similar metallicity.  Perhaps, Figure
\ref{fig:f18} gives an impression that the CNO sums for some wk Gb stars maybe greater than for
nearby giants. This is contrary to expectation for contamination of wk Gb atmospheres by CNO-cycled
material, casts doubt on the proposal that ON-cycled material is present in some wk Gb stellar
atmospheres,  and directs suspicion to systematic effects in the abundance analysis. This suspicion
should likely be directed at the C abundances which were derived from different spectral signature
for the two samples (see Section 4.1): the CH G-band for wk Gb stars and  C\,{\sc i} lines  and a
C$_2$ feature  for nearby giants.  A systematic error in the derived C abundance translates to an
error in the N abundance obtained from CN lines and, thence, into the CNO sum. A change of the CN
dissociation energy by 0.05\,eV (see section \ref{sect:nitrogen}) would result in a change of $<
0.1$ in the CNO sum. 

Other H-burning  processes occur in parallel with the CN-cycle and redistribution of the relevant
catalysts may lead to observable abundance anomalies. The following discussion is based on the NACRE
rates (see \nocite{1999A&A...347..572A}{Arnould}, {Goriely}, \&  {Jorissen} 1999).  The NeNa chain
provides a redistribution of Ne isotopes and an increase of the Na abundance and the MgAl chain
offers the possibility of changing the relative abundance of the three stable Mg isotopes:
$^{24}$Mg, $^{25}$Mg, and $^{26}$Mg.  Unfortunately, Ne and its isotopes  are not observable in K
giants but Na (see above) is observable.  Arnould et al. show that in a solar mix  Na increases by
about a factor of four  from running of the $^{22}$Ne($p,\gamma)^{23}$Na reaction over the
temperature range and proton consumptions levels required for efficient CN-cycling. Such a
quantitative estimate of the Na increase is dependent on several uncertain reaction rates in the
NeNa chain. The increase for a wk Gb star depends on the intrinsic $^{22}$Ne vs [Fe/H]  and [Na/H]
vs [Fe/H] relations. Figure \ref{fig:f11} shows that indeed the wk Gb stars are Na-rich relative to
the normal nearby giants.  This increase results from the NeNa chain operating in parallel with the
CNO-cycles. The MgAl chain can lead to the increase of $^{26}$Mg and a decrease of $^{25}$Mg
relative to the dominant isotope $^{24}$Mg. These alterations demand very deep mixing (i.e., hotter
temperatures than for activation of the NeNa chain) and a considerable conversion of H to He before
significant alteration of the Mg isotopic abundances. The inspection of our spectra indicates that
MgH lines are present in the wk Gb stars. Since the C$_2$ Swan lines that are often blended with key
MgH lines are not present in the wk Gb stars, an isotopic abundance determination is promising and
will be a future investigation. 

\section{Concluding remarks}\label{sect:conclusion}
Our abundance analysis of a complete sample of known northern wk Gb stars allows the illustration of
new and comprehensive results about their atmospheric composition including several results which
contradict published conclusions, often based on abundances for smaller samples of wk Gb stars. The
new results show that the atmospheres are seriously contaminated with products of the H-burning
CN-cycle and associated processes. Speculations about  episodes in the evolution of these stars
which result in the demarcation of wk Gb stars from normal giants are here curtailed. These
concluding remarks are limited to setting the stage on which the key distinguishing  episode in
stellar evolution must play.  Qualitative and certainly quantitative theoretical  exploration of
potential episodes are left to others.

A boundary condition generally accepted but not proven decisively from observations is that the
striking composition of a wk Gb star results from upheavals internal to the star and is not  imposed
by mass transfer  from or induced by an orbiting companion star
(\nocite{1984PASP...96..609T}{Tomkin}, {Sneden}, \&  {Cottrell} 1984;
\nocite{1992Obs...112..219G}{Griffin} 1992).  The binary fraction among wk Gb stars is thought to be
normal in sharp contrast, for example, to the Ba II K giants where all stars are binaries and  the
peculiar giant was created by transfer of $s$-process enriched material from its companion when it
was an AGB star \nocite{1983ApJ...268..264M}({McClure} 1983).  A comprehensive radial velocity
survey of wk Gb stars for orbital variations has yet to be undertaken.  Two stars -- HR\,1023 and
HR\,6791 -- are known to be spectroscopic binaries \nocite{1982A&A...105..318L, 1984PASP...96..609T,
1992Obs...112..219G}({Lucke} \& {Mayor} 1982; {Tomkin} {et~al.} 1984; {Griffin} 1992).\footnote{On
the assumption that the orbits are inclined at the average inclination for a random distribution of
orbits, these spectroscopic binaries according to the determined mass functions have masses
($M_1,M_2)$ in solar masses of ($4.4, 1.3$) for HR\,1023, and ($3.5,0.5$) for HR\,6791 where the
primary masses are taken from Table 3.} 

The wk Gb stars are systematically metal-poor relative to  local field giants (Figure
\ref{fig:f12}). This result shown here for the first time may possibly be, in part, a selection
effect. 
 
An important boundary condition is that the masses of wk Gb stars (2.5-5\Msun) are substantially
greater than the masses of  typical giants in the local field. This result,  first shown by
\nocite{2012A&A...538A..68P}{Palacios} {et~al.} (2012), arises from the rereduction of the
\textsl{Hipparcos} parallaxes \nocite{2007A&A...474..653V}({van Leeuwen} 2007).  This difference in
mass encourages the speculation (discussed above in Sect. \ref{sect:prelremarks}) that
rotationally-induced mixing in the main sequence progenitor may be the root cause of the abundance
anomalies of the wk Gb stars. Lower mass stars are rotationally-braked before reaching the main
sequence and observations show experience the mild effects of the standard first- dredge up, which
salts a low mass giant's atmosphere to a minor but observable degree with CN-cycled material, as
discussed by, for example, \nocite{2007AJ....133.2464L}{Luck} \& {Heiter} (2007). 

At the masses of the wk Gb stars, rapidly-rotating main sequence stars are common and rotational
braking not  a guaranteed phenomenon. Thus, one may speculate that particular combinations of
initial internal distribution of angular rotation rates,  rotationally-induced mixing and rotational
braking may lead to the severe contamination of the stellar interior by CN-cycled material. The
parameters for suitable combinations must be fairly stringent  or the wk Gb phenomenon would be the
norm and not a peculiar condition for G-K giants.  There is also a suspicion that the wk Gb giants
are a class apart with an unpopulated gap separating them from normal giants. This, if true for
giants of the same mass as the wk Gb stars, presumably translates to special constraints on origins
for wk Gb stars invoking rotation.   

In order to examine the `rotation' solution in more detail in advance of theoretical advances,
further observational investigation may be proposed. Three are sketched here. 

First, a comprehensive abundance analysis needs to be undertaken for a large sample of stars at the
masses of the wk Gb stars on the main sequence and, in particular in the Hertzsprung gap. Do wk Gb
stars make an appearance on the main sequence or in the gap before becoming giants? Note that, if
the spectroscopic temperatures are adopted, most of the wk Gb stars are in fact on the red side of
the gap (Figure \ref{fig:f2}).\footnote{\nocite{1994AJ....107.2211W}{Wallerstein} {et~al.} (1994)
determine Li abundances for 52 giants of 2-5\,$M_\odot$ in the gap and find no Li-rich examples.
Estimates of [N/C] from ultraviolet emission lines find no stars with N/C characteristic of wk Gb
stars.}

Second, there are several ways in which the present abundance analysis could be improved or
extended, apart from the obvious extension to wk Gb stars in the southern hemisphere beyond the
grasp of a Texas telescope. It is possible with due care given to correction for telluric lines to
obtain accurate measurements of C\,{\sc i} lines in the red (see above). In addition, infrared
spectra at 2.3 microns will provide measurements of the CO first-overtone bands which will yield a
different measure of the C abundance and, hence a check on the suggestion that N has been augmented
by partial ON-cycle conversion of some $^{16}$O to N. A more difficult observation would involve
spectra at 4.6 microns and a search for the isotopes $^{16}$O and $^{17}$O. 

Third, the search for wk Gb stars should be expanded. One  key question concerns the apparent clean
distinction between normal and wk Gb giants - do stars with intermediate C (and other) abundances
exist in detectable numbers? Is the wk Gb star phenomenon limited to particular metallicities?

\acknowledgements

We thank the referee for constructive comments and suggestions.
DLL thanks the Robert A. Welch Foundation of Houston, Texas for support through grant F-634.

\clearpage
\begin{landscape}
\begin{deluxetable}{lrrrrrrrrrrrrrrrrrrrrrrrr}
\tablecaption{Element abundances of the program stars with respect to H.\label{tab:abundances}}
\tablewidth{0pt}
\tablehead{
\colhead{Object} &\multicolumn{2}{c}{Na}     &\multicolumn{2}{c}{Mg}     &\multicolumn{2}{c}{Al}     &\multicolumn{2}{c}{Si}     &\multicolumn{2}{c}{Ca}     & \multicolumn{2}{c}{Ti} &
                  \multicolumn{2}{c}{Cr}     &\multicolumn{2}{c}{Fe}     &\multicolumn{2}{c}{Ni}     &\multicolumn{2}{c}{Y}      &\multicolumn{2}{c}{Eu} \\   
\colhead{}       &\colhead{sp.}&\colhead{ph.}&\colhead{sp.}&\colhead{ph.}&\colhead{sp.}&\colhead{ph.}&\colhead{sp.}&\colhead{ph.}&\colhead{sp.}&\colhead{ph.}& 
                  \colhead{sp.}&\colhead{ph.}&\colhead{sp.}&\colhead{ph.}&\colhead{sp.}&\colhead{ph.}&\colhead{sp.}&\colhead{ph.}&\colhead{sp.}&\colhead{ph.}&\colhead{sp.}&\colhead{ph.}
}
\startdata                                                                                                                                             
37\,Com             & 0.09&-0.11&     &     & 0.07&-0.11&-0.09& 0.05&-0.56&-0.82&-0.31&-0.67&-0.27&-0.57&-0.46&-0.50&-0.49&-0.54&-0.86&-1.15&     &      \\
BD\,+5$^{\circ}$593 & 0.17&-0.02&-0.10&-0.20&-0.05&-0.18&-0.12&-0.07&-0.12&-0.32&-0.12&-0.38&-0.22&-0.38&-0.23&-0.35&-0.27&-0.38&-0.31&-0.42&-0.20&-0.18 \\
HD\,18636           & 0.28& 0.17& 0.01&-0.07& 0.10& 0.01& 0.03& 0.06& 0.00&-0.14& 0.00&-0.20&-0.05&-0.17&-0.10&-0.17&-0.12&-0.21&-0.07&-0.15&-0.04&-0.02 \\
HD\,28932           & 0.04&-0.16&-0.15&-0.25&-0.06&-0.23&-0.19&-0.14&-0.20&-0.44&-0.16&-0.44&-0.12&-0.32&-0.30&-0.40&-0.37&-0.46&-0.32&-0.45&-0.27&-0.25 \\ 
HD\,31869           & 0.13& 0.09&-0.06&-0.08&-0.08&-0.11&-0.12&-0.12&-0.15&-0.19&-0.24&-0.29&-0.23&-0.27&-0.28&-0.31&-0.33&-0.35&-0.46&-0.48&     &      \\
HD\,40402           & 0.38& 0.27&-0.06&-0.11& 0.13& 0.04& 0.06& 0.11&-0.03&-0.16& 0.00&-0.16& 0.03&-0.10&-0.03&-0.08&-0.06&-0.10&-0.22&-0.33&     &      \\
HD\,49960           & 0.25& 0.19& 0.06& 0.03& 0.03&-0.02& 0.03& 0.05&-0.13&-0.20&-0.18&-0.27&-0.11&-0.16&-0.19&-0.22&-0.20&-0.22&-0.31&-0.35&-0.10&-0.10 \\
HD\,67728           & 0.44& 0.30&     &     &     &     &-0.21&-0.20&-0.21&-0.38&-0.37&-0.51&-0.06&-0.18&-0.23&-0.31&-0.26&-0.39&-0.36&-0.57&     &      \\
HD\,78146           & 0.38& 0.20& 0.08& 0.02& 0.09&-0.09& 0.12& 0.25&-0.04&-0.28&-0.15&-0.46& 0.04&-0.16&-0.08&-0.08&-0.03&-0.04&-0.20&-0.32&-0.11&-0.08 \\
HD\,82595           & 0.62& 0.49& 0.22& 0.15& 0.28& 0.17& 0.27& 0.31& 0.16& 0.01& 0.16&-0.04& 0.24& 0.10& 0.13& 0.08& 0.18& 0.12& 0.22& 0.19& 0.19& 0.20 \\
HD\,94956           & 0.27& 0.13&-0.13&-0.21& 0.06&-0.07& 0.01& 0.07&-0.13&-0.30&-0.12&-0.32&-0.03&-0.13&-0.15&-0.22&-0.19&-0.26&-0.20&-0.29&-0.13&-0.11 \\
HD\,120170          & 0.14&-0.03&-0.22&-0.31&-0.13&-0.26&-0.27&-0.24&-0.20&-0.39&-0.26&-0.51&-0.35&-0.50&-0.42&-0.51&-0.42&-0.52&-0.57&-0.67&     &      \\
HD\,120171          &-0.28&-0.29&-0.16&-0.17&-0.05&-0.07&-0.11&-0.10&-0.18&-0.20&-0.28&-0.31&-0.36&-0.38&-0.34&-0.34&-0.34&-0.35&-0.49&-0.50&     &      \\
HD\,132776          & 0.48& 0.30& 0.12& 0.08& 0.18& 0.01& 0.19& 0.33&-0.10&-0.24& 0.04&-0.27& 0.05&-0.13&-0.03&-0.02& 0.04& 0.05&-0.11&-0.23&-0.02& 0.01 \\
HD\,146116          & 0.15&-0.10&-0.12&-0.24&-0.09&-0.29&-0.13&-0.03&-0.20&-0.48&-0.23&-0.57&-0.25&-0.48&-0.32&-0.41&-0.32&-0.41&-0.40&-0.55&-0.33&-0.29 \\
HD\,188028          & 0.34& 0.08& 0.09&-0.04& 0.08&-0.14& 0.01& 0.14&-0.08&-0.38&-0.07&-0.48&-0.08&-0.37&-0.12&-0.23&-0.13&-0.22& 0.00&-0.05&-0.07&-0.03 \\
HD\,204046          & 0.38& 0.14& 0.35& 0.23& 0.16&-0.05& 0.08& 0.18& 0.02&-0.26&-0.02&-0.39& 0.03&-0.24&-0.03&-0.11&-0.04&-0.12&-0.06&-0.21& 0.02& 0.05 \\
HD\,207774          & 0.23& 0.04& 0.05&-0.07& 0.00&-0.08&-0.05& 0.02&-0.17&-0.40&-0.12&-0.43&-0.22&-0.42&-0.23&-0.33&-0.20&-0.31&-0.28&-0.32&-0.19&-0.17 \\ 
HR\,885             & 0.18& 0.14&-0.08&-0.11&-0.09&-0.13&-0.15&-0.14&-0.12&-0.17&-0.31&-0.39&-0.26&-0.29&-0.35&-0.38&-0.35&-0.39&-0.30&-0.28&     &-0.28 \\
HR\,1023            & 0.48& 0.28&     &     &     &     & 0.20& 0.10& 0.12&-0.14& 0.21&-0.16&     &     &-0.20&-0.41&-0.04&-0.29&     &     &     &      \\
HR\,1299            & 0.49& 0.45& 0.03& 0.01& 0.21& 0.18& 0.21& 0.23& 0.04& 0.00& 0.00&-0.06& 0.06& 0.02&-0.01&-0.02& 0.02& 0.02&-0.10&-0.12& 0.09& 0.08 \\
HR\,6757            & 0.41& 0.12&-0.03&-0.16& 0.12&-0.13& 0.07& 0.22& 0.09&-0.26& 0.00&-0.46& 0.09&-0.21&-0.04&-0.13&-0.06&-0.18& 0.05&-0.14&-0.04& 0.00 \\
HR\,6766            & 0.30& 0.19&-0.08&-0.14& 0.07&-0.03&-0.01& 0.03&-0.10&-0.24&-0.06&-0.23&-0.06&-0.16&-0.15&-0.21&-0.17&-0.23&-0.22&-0.29&-0.12&-0.10 \\
HR\,6791            & 0.56& 0.50& 0.06& 0.01& 0.19& 0.13& 0.21& 0.21& 0.05&-0.03& 0.03&-0.06& 0.14& 0.13& 0.04& 0.00& 0.04&-0.01&-0.10&-0.25& 0.03& 0.04 
\enddata
\end{deluxetable}
\clearpage
\end{landscape}

\end{document}